\documentclass[11pt,a4paper,pdfa=true]{article}
\pdfoutput=1
\usepackage{jcappub}
\usepackage{color}
\usepackage[normalem]{ulem}

\usepackage[T1]{fontenc}
\usepackage{graphicx,amssymb,amsmath,wasysym}

\begin{document}
\hfill {\tt CFTP/14-019} 
	
\hfill {\tt TUM-HEP 970/14}
	
\hfill {\tt IFIC/14-78}

\title{Future sensitivity of neutrino telescopes to dark matter
  annihilations from the cosmic diffuse neutrino signal} 

\author[a]{\'Angeles Molin\'e}
\author[b]{, Alejandro Ibarra}
\author[c]{and Sergio Palomares-Ruiz}
 
\affiliation[a]{CFTP, Instituto Superior Tecnico, Universidade Tecnica de Lisboa, \\ 
	Av. Rovisco Pais 1, 1049-001 Lisboa, Portugal} 
\affiliation[b]{Physik-Department T30d, Technische Univertit\"at M\"unchen, \\ James-Franck-Stra\ss{}e, 85748 Garching, Germany} 
\affiliation[c]{Instituto de F\'isica Corpuscular (IFIC), CSIC-Universitat de Val\`encia, \\  
	Apartado de Correos 22085, E-46071 Val\`encia, Spain}

\abstract{
Cosmological observations and cold dark matter N-body simulations indicate that our Universe is populated by numerous halos, where dark matter particles annihilate, potentially producing Standard Model particles.  In this paper we calculate the contribution to the diffuse neutrino background from dark matter annihilations in halos at all redshifts and we estimate the future sensitivity to the annihilation cross section of neutrino telescopes such as IceCube or ANTARES.  We consider various parametrizations to describe the internal halo properties and for the halo mass function in order to bracket the theoretical uncertainty in the limits from the modeling of the cosmological annihilation flux.  We find that observations of the cosmic diffuse neutrino flux at large angular distances from the galactic center lead to constraints on the dark matter annihilation cross section which are complementary to (and for some extrapolations of the astrophysical parameters, better than) those stemming from observations of the Milky Way halo, especially for neutrino telescopes not pointing directly to the Milky Way center, as is the case of IceCube.
}


\maketitle

\section{Introduction}
\label{sec:intro}

Cosmological observations have demonstrated that approximately 80\% of the matter content of the Universe is comprised by a new particle (or particles) not contained in the Standard Model (SM), the dark matter (DM) particle. Despite the overwhelming evidence for the existence of this new particle, its properties are still largely unconstrained by observations (for reviews, see Refs.~\cite{Jungman:1995df, Bergstrom:2000pn, Munoz:2003gx, Bertone:2004pz, Bertone:2010}).  In fact, many particle physics candidates have been proposed in the literature, with very disparate masses and interaction cross sections with ordinary matter, and shown to have an abundance today compatible with the Planck determination of the cold DM density $\Omega_{\textrm{c}} h^2=0.1199\pm 0.0027$~\cite{Ade:2013zuv}.    

Among the myriad of DM candidates proposed, weakly interacting massive particles (WIMPs) stand as one of the most plausible DM candidates. WIMPs are assumed to be stable in cosmological time scales and  to interact in pairs with SM particles.  At very high temperatures, this interaction keeps WIMPs in thermal equilibrium with the SM particles. In particular, when the temperature of the Universe drops below the WIMP mass, the WIMP number density per comoving volume exponentially decreases.  This catastrophic depletion of DM particles continues until the expansion rate of the Universe becomes larger than the annihilation rate.  Below this temperature, and due to the expansion of the Universe, the average distance between WIMPs becomes so large that annihilations occur very rarely and, therefore, the WIMP number density per comoving volume remains practically constant until today, i.e., it freezes-out, constituting the DM population we observe in our Universe. 

Immediately after the WIMP freeze-out, the DM distribution was roughly homogeneous and isotropic throughout the whole Universe, except for small density fluctuations generated by the inflaton field~\cite{Hawking:1982cz, Starobinsky:1982ee, Guth:1982ec, Bardeen:1983qw}.  Numerical N-body simulations show that the overdensity regions accrete the DM located in their surroundings and seed the formation of larger and larger structures (see, e.g., Refs.~\cite{Davis:1985rj, Springel:2005nw, BoylanKolchin:2009nc, Klypin:2010qw}).  The baryonic matter, also distributed roughly homogeneously and isotropically at very early times, followed the gravitational potential wells generated by DM, forming structures which are identified with galaxies and clusters of galaxies. In these regions, the density of DM particles is orders of magnitude larger than the average matter density and their annihilations open the possibility of detecting the DM particles via the identification of their annihilation products over the expected backgrounds~\cite{Bergstrom:2001jj, Ullio:2002pj, Taylor:2002zd, Beacom:2006tt}.  

In this paper we focus on the possibility of detecting the neutrinos which are produced in WIMP annihilations. The closest region to the Earth where an overdensity of WIMPs is generally expected, and which constitutes a prime target for DM detection, is the Sun~\cite{Silk:1985ax, Srednicki:1986vj, Hagelin:1986gv, Gaisser:1986ha, Kamionkowski:1991nj, Bottino:1991dy, Halzen:1991kh, Bergstrom:1996kp, Barger:2001ur, Cirelli:2005gh, Mena:2007ty, Blennow:2007tw, Wikstrom:2009kw, Kappl:2011kz, Rott:2011fh, Das:2011yr, Bernal:2012qh, Rott:2012qb, Ibarra:2013eba, Baratella:2013fya, Ibarra:2014vya}. DM WIMPs scatter with the matter in the solar interior, lose energy and eventually sink to the solar core.  The number of DM particles captured inside the Sun depends on their scattering cross section with nucleons, hence the non-observation of a high energy neutrino flux correlated to the direction of the Sun allows setting upper limits on this quantity. Searches have been conducted by the SuperKamiokande~\cite{Tanaka:2011uf} (see also Refs.~\cite{Kappl:2011kz, Bernal:2012qh, Rott:2012qb}), IceCube~\cite{Aartsen:2012kia}, Baksan~\cite{Boliev:2013ai} and ANTARES~\cite{Adrian-Martinez:2013ayv} collaborations, with null results.  Remarkably, for annihilation channels producing a hard neutrino spectrum, such as $W^+W^-$ or $\tau^+\tau^-$, and at low masses for searches with MeV neutrinos~\cite{Bernal:2012qh, Rott:2012qb}, the limits on the spin-dependent interaction cross section are more stringent than those reported by the most sensitive current direct detection experiments probing the same interaction, SIMPLE~\cite{Felizardo:2011uw}, PICASSO~\cite{Archambault:2012pm} and COUPP~\cite{Behnke:2012ys}.  Other stars in the Milky Way are also sources of high energy neutrinos, however their large distance to the Earth makes their detection very challenging in practice.

Another prime target for DM detection is the Milky Way center. Despite being located at about 8.5 kpc from the Earth, this is the region of the galaxy with the highest DM density.  The large neutrino luminosity can then compensate the large distance between the galactic center and the Earth, representing a potentially observable signal~\cite{Yuksel:2007ac, PalomaresRuiz:2007eu, PalomaresRuiz:2007ry, Liu:2008ci, Erkoca:2009by, Buckley:2009kw, Liu:2009ac, Mandal:2009yk, Covi:2009xn, Erkoca:2010qx, Erkoca:2010vk, Lee:2011nt, Esmaili:2012us, Lee:2012pz, Avrorin:2014vca}.  The non-observation of a significant excess of high-energy neutrino events in the direction of the galactic center with respect to the expected backgrounds allows setting rather stringent limits on the annihilation cross section of DM particles with masses in the range 100 GeV-10~TeV~\cite{Abbasi:2012ws, Aartsen:2013mla, Hernandez-Rey:2014ssa, Zornoza:2014cra, Wendell:2014dka}.  At even larger distances from the Earth, the next targets for neutrino detection are the Milky Way satellites, other galaxies in our local cluster and galaxy clusters~\cite{Sandick:2009bi, Yuan:2010gn, Dasgupta:2012bd, Murase:2012rd}.  A search for the neutrinos produced in DM annihilations in several dwarf galaxies, the Andromeda galaxy, as well as in the Coma and Virgo clusters was conducted in Ref.~\cite{Aartsen:2013dxa}, resulting in limits on the annihilation cross section which are complementary to the limits from the observations of the galactic center.  Finally, high energy neutrinos could be searched for in the diffuse extragalactic background~\cite{Beacom:2006tt, Murase:2012xs}, which is the goal of this work.  In particular, we focus on searches with km$^3$ \v{C}erenkov neutrino telescopes like ANTARES and IceCube. 

In the theoretical paradigm of hierarchical structure formation, initial gravitational instabilities constitute the seeds for the formation of cold DM (CDM) halos, which dragged baryons into their potential wells where galaxies and clusters of galaxies formed. Numerical CDM N-body simulations have proven to be crucial for understanding structure formation in the Universe, both to study structural properties of individual halos and to determine the abundance of halos as a function of time and mass.  These simulations reveal the existence of a population of halos, even at high redshifts, where DM annihilations took place.  The neutrino luminosity from a single halo depends on the total mass of the halo and on the distribution of DM particles within it.  The combined neutrino emission from all halos at all redshifts produces a nearly isotropic neutrino flux that could be detected at the Earth over the atmospheric neutrino background~\cite{Beacom:2006tt, Murase:2012xs}.  Nevertheless, these simulations only cover a limited range in halo masses and redshifts, so the calculation of the neutrino flux relies on extrapolations, which could incur large uncertainties. Despite this, for neutrino telescopes not directly observing the galactic center, as is the case of IceCube, the sensitivity to DM annihilations from observations of the cosmic diffuse neutrino signal could be competitive with the one from observations of the galactic center.  In this paper we estimate the future sensitivity to the DM annihilation cross section from the cosmic diffuse neutrino signal with km$^3$ neutrino telescopes, with emphasis on the astrophysical uncertainties on the limits. The paper is organized as follows. In Section~\ref{sec:nuflux} we describe the calculation of the isotropic neutrino flux and discuss the various relevant uncertainties. In order to approximately bracket these uncertainties and the differences obtained with various simulations and extrapolations, we consider several parametrizations for the concentration parameter of individual halos and for the halo mass function.  We also consider the neutrino flux from the Milky Way and compare it with the cosmic diffuse neutrino signal.  In Section~\ref{sec:event} we describe the two type of events in neutrino telescopes we consider in this work: through-going muon events and showers.  Using a simplified modeling of the detector, in Section~\ref{sec:bounds} we discuss the sensitivity to the DM annihilation cross section a km$^3$ neutrino telescope could achieve after 10~years of data-taking. Finally, in Section~\ref{sec:conclusions} we draw our conclusions.

\section{Neutrino fluxes}
\label{sec:nuflux}

Numerical CDM N-body simulations have demonstrated that the density fluctuations in the distribution of DM particles in the very early Universe seed the formation of DM halos, where the annihilation rate can be enhanced and which therefore constitute prime targets for DM detection. The density distribution in one individual halo is commonly described by the Navarro-Frenk-White (NFW) profile~\cite{Navarro:1995iw, Navarro:1996gj}:
\begin{equation}
\label{eq:nfw1}
\rho(r)=\frac{4\,\rho_{\rm s}}{r/r_{\rm s}\left(1+r/r_{\rm s}\right)^2} ~,
\end{equation}
where $r$ is the radial coordinate, $r_{\rm s}$ is the scale radius, defined as the distance to the center of the halo for which the logarithmic slope of the profile satisfies $\frac{d \log\rho}{d \log\,r}\Big|_{r_s}=-2$, and $\rho_{\rm s}$ is the DM density at the scale radius. This profile is completely specified by two independent parameters.  One can use $\rho_{\rm s}$ and $r_{\rm s}$, or for instance, use the total halo mass $M$ and the concentration parameter $c$, which we define below.  It should be borne in mind that recent works~\cite{Navarro:2003ew, Merritt:2005xc, Graham:2005xx, Stadel:2008pn, Navarro:2008kc, Reed:2010gh, Dutton:2014xda} have shown that a spherical Einasto profile~\cite{Einasto:1965}, with three independent parameters instead of two, provides a better description of spherical CDM halos than the NFW profile~\cite{Navarro:1995iw, Navarro:1996gj}.  Nevertheless, since most studies of the evolution of structural parameters for CDM halos use a NFW fit, this is the choice we adopt in our analysis.   

There are two main approaches to identify DM halos in N-body simulations: through the friends-of-friends (FOF) finder~\cite{Davis:1985rj} or through the spherical overdensity (SO)   algorithm~\cite{Lacey:1994su}.  With the FOF finder, the particles in the simulation are linked together if the distance to the nearest group member is smaller than a fraction of the mean distance between particles (the linking length).  On the other hand, the SO algorithm assumes the shape of the halos to be spherical\footnote{Although this can be corrected to  describe non-spherical halos (see Ref.~\cite{Zemp:2011ed} for a critical discussion of different methods), which has implications in DM searches~\cite{Zemp:2008gw, Pato:2010yq, Bernal:2014mmt}.}, with some differences among the various SO approaches stemming from the way the center of the halo is determined.  The FOF algorithm has the advantage that the halo shape is not fixed a priori, however, it can spuriously link halos by a bridge of particles.  In our analysis we consider results obtained with the SO algorithm and with different halofinding codes.  In this approach, a DM halo at redshift $z$ is characterized by one parameter, the overdensity $\Delta (z)$ with respect to the critical density, $\rho_{\rm c}(z)$.  Spheres are grown from the center until the enclosed density of the sphere is equal to $\Delta (z)$. Then, a sphere of radius $r_\Delta$ around the halo center contains a mass  
\begin{equation}
\label{eq:D}
M  = \frac{4\pi}{3} \, \Delta(z) \, \rho_{\rm c}(z)\,r_{\Delta}^{3} ~.
\end{equation}

Various choices of the overdensity parameter have been considered in the literature.  Halos are usually defined either with a fixed value for the overdensity, normally $\Delta (z) = 200$, or using the virial overdensity based on spherical collapse, which depends on the details of the cosmology model as well as on the redshift~\cite{Bryan:1997dn},    
\begin{equation}
\label{eq:dvir}
\Delta(z) = \Delta_{\rm vir} (z) =  18 \pi^2 + 82 \, (\Omega_{\rm
  m}(z)-1) - 39 \, (\Omega_{\rm m}(z)-1)^2 ~,  
\end{equation}
where $\Omega_{\rm m}(z)=(1+\Omega_{\Lambda}/\Omega_{\rm m,0}(1+z)^{3})^{-1}$, while $\Omega_{\rm m,0}$ and $\Omega_{\Lambda}$ are the matter and cosmological constant densities at present time normalized to the critical density $\rho_{c,0} = 1.05 \, h^2 \times 10^{-5} \, {\rm GeV}\,{\rm cm}^{-3}$. Note that when adopting $\Delta(z)=200$ for the overdensity, the time dependence of the halo mass follows only from $\rho_{\rm c}(z)$, in contrast to the choice
$\Delta_{\rm vir}(z)$.   

The annihilations of DM into SM particles occurring in these overdensity regions generate a neutrino flux that could be detected at the Earth.  To calculate the total cosmological diffuse neutrino flux we follow the halo model approach (alternatively, one can also consider the power spectrum approach~\cite{Serpico:2011in, Sefusatti:2014vha}).  We first consider the contribution from a single halo located at the comoving distance $\chi(z)$ with total mass $M$ and concentration parameter $c \equiv r_\Delta/r_s$, which is defined as the ratio of the radius of the halo, $r_\Delta$, and the scale radius, $r_s$. Disregarding for the moment the effect of neutrino oscillations, the differential flux of neutrinos with  flavor $\beta$ reads:   
\begin{equation}
\frac{d
  \phi_{\nu_\beta}(E_0)}{dE_{0}}\Big|_{M,c;z} = \frac{1}{4\pi \chi^2(z)}
\frac{\langle\sigma \upsilon\rangle}{2 \, m^2_{\rm DM}} 
 \int 4\pi r^2 \rho^2(r;M,c) \, dr \, \sum_{i} {\rm Br}_{i}
 \frac{dN_{\nu_\beta, i}(E)}{dE}   ~,    
\end{equation}
where $\langle\sigma \upsilon\rangle$ is the averaged DM annihilation cross section times relative velocity in the pair, $m_{\rm DM}$ the DM mass, and $dN_{\nu_\beta,i}/dE$ is the differential energy spectrum of neutrinos of flavor $\beta$ per annihilations into the channel $i$, with branching ratio ${\rm   Br}_{i}$. In this formula, $E_0$ is the energy received at the Earth and $E=E_0 (1+z)$ is the energy at the source. To calculate the spectrum at the source, we make use of the results presented in Ref.~\cite{Cirelli:2010xx}, which were computed using PYTHIA 8.1~\cite{Sjostrand:2007gs} and include electroweak corrections~\cite{Ciafaloni:2010ti}.  We also assume DM annihilations with 100\% branching fraction into some exemplary channels, in order to cover a wide class of DM models.

The differential flux can also be cast as:
\begin{equation}
\frac{d
  \phi_{\nu_\beta}(E_0)}{dE_{0}}\Big|_{M,c;z} = \frac{1}{4\pi \chi^2(z)} \frac{\langle\sigma \upsilon\rangle}{2} \, \frac{\rho_{\rm m,0}^2}{m^2_{\rm DM}} \, \frac{\Delta(z) \, \rho_{\rm c}(z)}{\rho_{\rm m,0}} \, \frac{M}{\rho_{\rm m,0}} \, \xi^2_{\rm M}(M,c;z) \, \sum_{i} {\rm Br}_{i} \frac{dN_{\nu_\beta, i}(E)}{dE}   ~,
\end{equation}
where $\xi^2_{\rm M}(M,c;z)$ is given by~\cite{Taylor:2002zd} 
\begin{equation}
\label{eq:xiM2}
\xi^2_{\rm M}(M,c;z) = \frac{M}{\Delta(z) \, \rho_{\rm c} (z)} \,
\frac{\int 4 \pi r^2 \rho^2(r;M,c) \, dr}{(\int 4 \pi r^2 \rho(r;M,c) \, dr)^2} ~,   
\end{equation}
and measures the enhancement in the flux due to the overdensity of DM particles compared to the one produced by a smooth DM distribution with density $\Delta(z) \, \rho_{\rm c}(z)$, in a volume $4\pi r_\Delta^3/3$ as defined in Eq.~(\ref{eq:D}).  For a NFW profile, the enhancement of the signal of a single halo, Eq.~(\ref{eq:xiM2}), has an analytical solution, which in terms of the concentration parameter $c=r_\Delta/r_s$ reads: 
\begin{equation}
\label{eq:xiM2NFW}
\xi^2_{\rm M}(M,c;z)\Big|_{\rm NFW} = \frac{1}{9} \,
\frac{{c}^{3} \left(1-(1+c)^{-3}\right)}{\left[ \ln(1+c)-c (1+c)^{-1} \right]^{2}} ~.    
\end{equation}

In a spherical shell with comoving distance between $\chi$ and $\chi+d\chi$ there exists a population of halos with different masses and concentration parameters.  The comoving number of halos in the interval of mass $[M, M+dM]$ is parametrized by the halo mass function $dn(M,z)/dM$.  In order to compute the cosmological signal from DM annihilations it is then necessary to include the contribution from all halos at all redshifts.  On the other hand, most numerical N-body simulations explore rather narrow mass and redshift ranges, $M\sim (10^{10} - 10^{15}) \, h^{-1} M_\odot$ and $z<2-5$, thus incurring large uncertainties in the predictions of the neutrino flux from the extrapolations, both in mass and redshift, of the halo mass function inferred from simulations.  In particular, it is necessary to include the contribution from the smallest halos, which are presumably very copious in our Universe.  It has been shown that the free-streaming of DM particles from high to low density regions~\cite{Green:2005fa} and the effect of acoustic oscillations~\cite{Loeb:2005pm, Bertschinger:2006nq} generate an exponential cutoff in the power spectrum, which translates into a minimum halo mass.  Nevertheless, and due to the dependence of these processes on the particle physics and cosmological models~\cite{Bringmann:2009vf, Cornell:2012tb, Gondolo:2012vh}, the value of the minimum halo mass is poorly constrained and lies in the wide range $M_{\rm min}=10^{-11}-10^{-4}M_\odot$; in our analysis we set the minimum mass to $M_{\rm min} = 10^{-6} \, M_\odot$ (see, e.g., Ref.~\cite{SiegalGaskins:2008ge, Ng:2013xha, Sanchez-Conde:2013yxa, Ackermann:2015tah} for recent discussions about the impact of this parameter).  Various choices for the halo mass function presented in the literature are compiled in Subsection \ref{subsec:massfunction}. Furthermore, the distribution of concentration parameters, $P(c)$, is found to follow a log-normal distribution~\cite{Jing:1998xj} (see, however, Refs.~\cite{Reed:2010gh, Bhattacharya:2011vr, Diemer:2014gba}),
\begin{equation}
\label{eq:P}
P(c) = \frac{1}{c \, \ln 10 \, \sqrt{2 \pi} \, \sigma_{\log_{10} c_\Delta}} \,  
e^{-\frac{(\log_{10} c-\log_{10} c_\Delta)^2}{2 \, \sigma_{\log_{10} c_\Delta}^2}} ~,
\end{equation}
where $c_\Delta(M,z)$ is the center value of the concentration parameter for a DM halo with mass $M$ at redshift $z$, which is discussed in Subsection~\ref{subsec:concentration}.  For the dispersion of the concentration parameter we adopt $\sigma_{\log_{10} c_\Delta} = 0.2$~\cite{Jing:1998xj, Bullock:1999he}. Note that this implies an enhancement factor slightly larger, a few tens of percent, when other smaller values~\cite{Maccio':2006nu, Neto:2007vq, Maccio':2008xb, Dutton:2014xda} are considered.   

Then, the neutrino flux originated from DM annihilations in halos in a comoving volume at the comoving distance $\chi(z)$ is 
\begin{eqnarray}
\frac{d \phi_{\nu_\beta}(E_0)}{dE_{0}}\Big|_{\chi(z)} & = & \frac{1}{4\pi \chi^2(z)} \frac{\langle\sigma  \upsilon\rangle}{2} \, \frac{\rho_{\rm m,0}^2}{m^2_{\rm DM}} \,  
 \int_{M_{\rm min}} dM \frac{dn(M,z)}{dM} \times \nonumber \\
 & & \int dc \, P(c) \, \frac{\Delta(z) \, \rho_c(z)}{\rho_{\rm m,0}} \, \frac{M}{\rho_{\rm m,0}} \, \xi^2_{\rm M}(M,c;z) \, \sum_{i} {\rm Br}_{i} \frac{dN_{\nu_\beta,i}(E)}{dE}   ~.
\end{eqnarray}
This expression can be cast as
\begin{equation}
\frac{d \phi_{\nu_\beta}(E_0)}{dE_{0}}\Big|_{\chi(z)}=\frac{1}{4\pi \chi^2(z)} \frac{\langle\sigma  \upsilon\rangle}{2} \,
\frac{\rho^2_{\rm m,0}}{m^2_{\rm DM}} \, \xi^2(z)  \, \sum_{i} {\rm Br}_{i}
\frac{dN_{\nu_\beta, i}(E)}{dE}   ~,    
\end{equation}
where the enhancement factor over all halos at a given redshift is given by~\cite{Taylor:2002zd}  
\begin{equation}
\label{eq:xi2}
\xi^2(z)= \frac{\Delta(z) \, \rho_{\rm c}(z)}{\rho_{\rm m,0}} \,
\int_{M_{\rm min}} dM \frac{M}{\rho_{{\rm m},0}}\frac{dn(M,z)}{dM}\int
dc \, P(c) \, \xi^2_{\rm M}(M,c;z) ~. 
\end{equation}
The contribution from all halos is obtained integrating over all comoving distances. The result is~\cite{Ullio:2002pj, Taylor:2002zd} 
\begin{equation}
\label{eq:flux}
\frac{d \phi_{\nu_\beta}}{dE_{0}} = \frac{\langle\sigma \upsilon\rangle}{2} \frac{\rho^{2}_{\rm m,0}}{m_{\rm DM}^{2}} \, \int \frac{dz}{H(z)} \, \xi^{2}(z) \, \sum_{i} {\rm Br}_{i}
\frac{dN_{\nu_\beta, i}(E)}{dE}   ~,    
\end{equation}
where we have integrated up to $z_{\rm max}=20$ and we have used $d\chi=dz/H(z)$, with $H(z)= H_0 \, \sqrt{\Omega_{\rm m,0}(1+z)^{3} +  \Omega_{\Lambda}} \equiv H_0 \, h(z)$ the Hubble parameter as a function of redshift.  Here, $H_0= 100 \,h \, {\rm km}\,{\rm s}^{-1}\,{\rm Mpc}^{-1}$ is the Hubble parameter at the present time (we have set the contribution from curvature to zero, $\Omega_k=0$).  

So far we have neglected the neutrino flavor conversion during propagation due to the leptonic mixing.  Over cosmic distances, the neutrino mass eigenstates undergo many cycles of flavor oscillations. Hence, the effect of oscillations averages out and the neutrino flavor at the Earth is practically independent of the distance traveled. Then, the flux of neutrinos of flavor $\alpha$ that reaches the Earth, including the effect of oscillations, finally reads:  
\begin{equation}
\label{eq:prob}
\frac{d\phi_{\nu_\alpha}}{dE_0}\Big|_{\rm w/\,osc.} = \sum_{\beta,i}
|U_{\alpha,i}|^2 \, |U_{\beta,i}|^2 \, \frac{d\phi_{\nu_\beta}}{dE_0}\Big|_{\rm w/o\,osc.} ~, 
\end{equation}
where $U$ is the leptonic mixing matrix and the flux $d\phi_{\nu_\beta}/dE_0\Big|_{\rm w/o\,osc.}$, without including neutrino oscillations, is given by Eq.~(\ref{eq:flux}).  For the numerical analysis we use the latest $\nu$\textit{fit} results~\cite{GonzalezGarcia:2012sz} (see also Refs.~\cite{Tortola:2012te, Fogli:2012ua}).

\subsection{Concentration parameter}
\label{subsec:concentration}

There are basically two ways to determine the concentration parameter from N-body simulations~\cite{Navarro:1995iw, Navarro:1996gj, Jing:1998xj, Bullock:1999he, Eke:2000av, Wechsler:2001cs, Zhao:2002rm, Maccio':2006nu, Neto:2007vq, Gao:2007gh, Duffy:2008pz, Maccio':2008xb, Zhao:2008wd, Klypin:2010qw, MunozCuartas:2010ig, Reed:2010gh, Prada:2011jf, Bhattacharya:2011vr, Ludlow:2013bd, Ludlow:2013vxa, Dutton:2014xda, Diemer:2014gba}.  On one side, one can fit the density profiles of the resolved halos of radius $r_\Delta$ with a given profile, obtain $r_s$ and from there, define $c \equiv r_\Delta/r_s$.  On the other hand, instead of the mass, one can use the maximum of the circular speed, which does not depend on the density profile, and relate it to a particular density profile in order to extract $c$~\cite{Klypin:2010qw, Prada:2011jf}.  Although, these methods give very similar results at low redshifts, the latter systematically predicts $\sim$15\% larger concentrations at high masses and redshifts~\cite{Prada:2011jf, Dutton:2014xda}.  Moreover, it has been pointed out that it could be affected by transient features which would tend to overestimate $c$~\cite{Ludlow:2012wh}. 

Current analyses of the halo concentration show that, within the limited mass and redshift ranges of simulations, $\sim 10^{10} - 10^{15} \, h^{-1} M_\odot$ and $z<2-5$, the concentration parameter decreases with increasing halo mass and redshift.  Beyond these ranges, several extrapolation procedures have been proposed in the literature. Power-law extrapolations for the concentration--mass relation are common. However, it has been recently found that the smallest halos have a steep inner slope~\cite{Ishiyama:2010es, Anderhalden:2013wd, Ishiyama:2014uoa} (the central regions could not be resolved in earlier simulations~\cite{Diemand:2005vz}) and that the concentration parameter has little dependence on the halo mass for masses in the range $10^{-6} - 10^{-4} \, M_\odot$~\cite{Ishiyama:2014uoa}. In order to approximately bracket the uncertainties and differences obtained with various computations and extrapolations, in this work we consider two parametrizations for the concentration--mass relation which present a flattening of the concentration parameter for low-mass halos and another one with a power-law behavior. Let us stress again that extrapolating simple power laws beyond the validity of the simulations is highly disfavored by the recent results on microhalos.

\begin{table}[t]
\begin{center}
\begin{tabular}{| l l l l l l l l | }
\hline
$c_\Delta(M, z)$ & $\Omega_{\rm m,0}$ & $\Omega_\Lambda$ &
$\Omega_{\rm b,0}$ & $h$ & $n_s$ & $\sigma_{8}$ & $\Delta$ \\ \hline
\hline 
P12~\cite{Prada:2011jf} & 0.27 & 0.73 & 0.0469 & 0.7 & 0.95 & 0.82 & 200 \\ \hline
B01~\cite{Bullock:1999he, Dutton:2014xda} & 0.3175 & 0.6825 & 0.0490 & 0.671 & 0.9624 & 0.8344 & $\Delta_{\rm vir}$ \\ \hline
DM14~\cite{Dutton:2014xda} & 0.3175 & 0.6825 & 0.0490 & 0.671 & 0.9624 & 0.8344 & 200 \\ \hline
\end{tabular}
\end{center}
\caption{Values of the parameters used in the simulations we consider to parametrize the concentration parameter $c_\Delta(M,z)$.  Note that for B01 we use a modified $K$ parameter, which best matches Planck cosmology~\cite{Dutton:2014xda}.}
\label{tab:conc}
\end{table}

\begin{figure}[t]
\includegraphics[width=\textwidth]{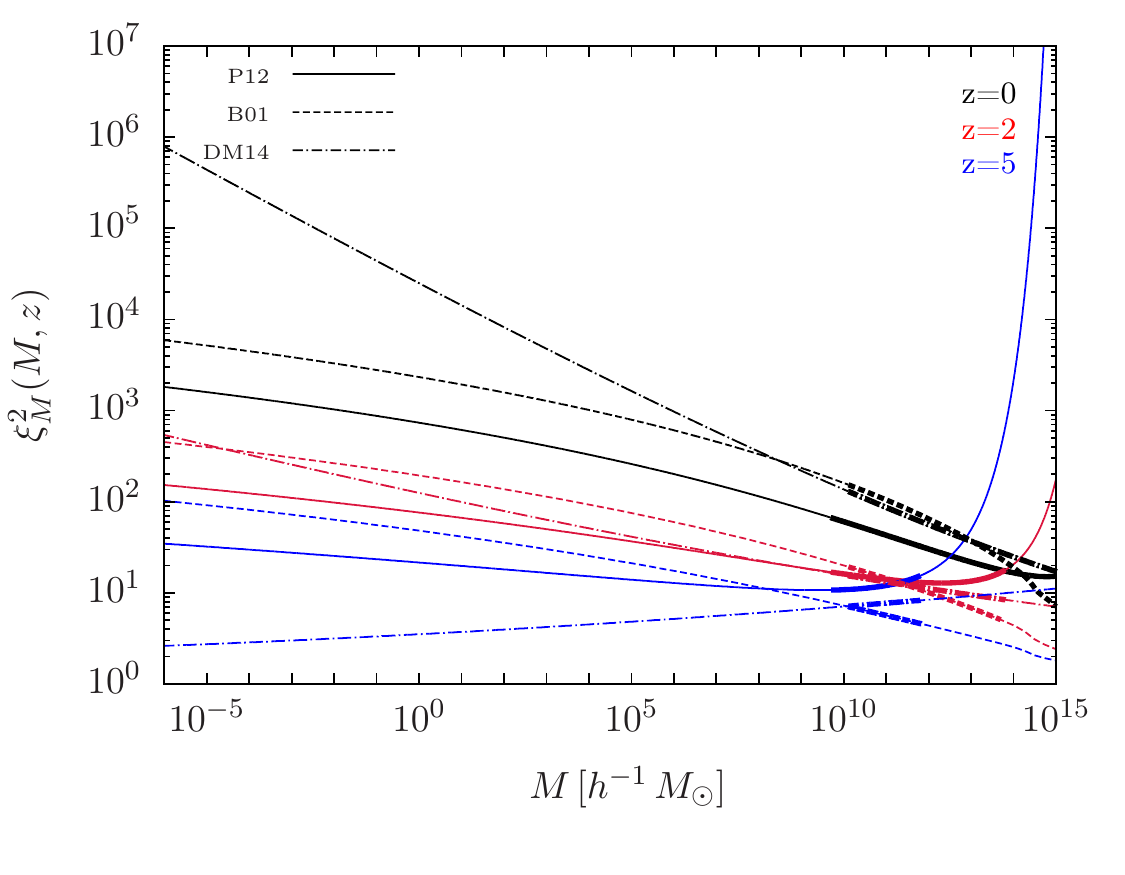}
\caption{The enhancement factor $\xi^2_{\rm M} (M,z)$ for individual halos with mass $M$ (in the range of masses we consider) and for three values of the redshift, from top to bottom $z=0, 2, 5$ (black, red and blue lines, respectively), for three parametrizations of the concentration parameter $c_\Delta(M,z)$, with parameters indicated in Tab.~\ref{tab:conc}: P12 (solid lines), B01 (dashed lines) and DM14 (dotted lines). On the high-mass regime, thinner lines are used for halos with masses above the maximum mass shown in Fig.~\ref{fig:dndM} for each case and redshift, i.e., halos with a very low probability to have been formed at that redshift. On the low-mass regime, thinner lines are used to indicate regions outside the range probed by the simulations.
} 
\label{fig:xiM2}
\end{figure}

We use the results for $c_{200}$ obtained with the MultiDark simulation (P12)~\cite{Prada:2011jf}, valid in the range $\sim 10^{11} h^{-1} M_\odot - 10^{15} \, h^{-1} M_\odot$ at $z=0$ and based on the maximum circular speed method.  They confirm the flattening and upturn in the concentration parameter at high masses and redshifts found a few years ago~\cite{Klypin:2010qw}, although this has been claimed to be due to unrelaxed halos in transient stages of their evolution~\cite{Ludlow:2012wh}, and show that the concentration parameter can be expressed in terms of $\sigma(M,z)$, the root mean square (rms) density fluctuation at redshift $z$, in linear theory, in spheres containing a mean mass $M$. In this way, the dependence on the cosmological parameters would be automatically included, simplifying the comparison with other works, which cannot be scaled when using approximations in terms of halo mass and redshift.  However, they found this relation not to be universal, but changing with time~\cite{Prada:2011jf, Dutton:2014xda}, so an extra redshift dependence was added, in disagreement with Ref.~\cite{Ludlow:2013vxa}.  A positive feature of this parametrization is that when applying it to the very low end of the mass range, no extrapolation is being performed.  Indeed, it has been shown to work well at halo masses as low as $10^{-6} \, h^{-1} M_\odot$~\cite{Sanchez-Conde:2013yxa} by contrasting it with recent results on microhalo properties~\cite{Ishiyama:2014uoa}.  

We also consider the Bullock {\sl et al.} model (B01)~\cite{Bullock:1999he} but with a modified $K$ parameter~\cite{Maccio':2008xb} that provides the best-fit for the Planck cosmology, $K=4.2$~\cite{Dutton:2014xda}.  This parametrization is obtained for $c_{\rm vir}$ and its time evolution is inversely proportional to $(1+z)$.  

Finally, we also consider one case (DM14) with results from simulations using the method based on fitting the density profile~\cite{Dutton:2014xda}.  The results are valid in the mass range $\sim 10^{10} - 10^{15} \, h^{-1} M_\odot$ and up to $z=5$~\cite{Dutton:2014xda}.  The evolution of the concentration--mass relation is well fitted by a power-law (see also, e.g., Refs.~\cite{Maccio':2006nu, Neto:2007vq, Maccio':2008xb, MunozCuartas:2010ig}),
\begin{equation}
\log_{10} c_\Delta = a(z) \log_{10} \,(M /[h^{-1} M_{\odot} ]) + b(z) ~,
\end{equation}
and we use the results of the fits for $c_{200}$. Let us also note that DM14 did not confirm the upturn in the evolution of the concentration parameter at high redshifts and masses found in other works such as P12~\cite{Klypin:2010qw, Prada:2011jf}.    

The cosmological parameters adopted in the three simulations we consider are given in Tab.~\ref{tab:conc}.  In Fig.~\ref{fig:xiM2}, using the NFW parametrization, we show the single halo enhancement $\xi^2_{\rm M}$, Eq.~(\ref{eq:xiM2NFW}), as a function of the halo mass $M$ (in the range of masses we consider) and for three values of the redshift, $z=0, 2, 5$, for the three parametrizations we discuss.

\subsection{The halo mass function}
\label{subsec:massfunction}

The number of halos per unit mass as a function of mass and redshift is parametrized by the comoving halo mass function~\cite{Jenkins:2000bv},    
\begin{equation}
\label{eq:dndm}
\frac{dn(M,z)}{dM} = \frac{\rho_{\rm m,0}}{M^{2}} \,
\frac{d\,\mbox{ln}\sigma^{-1}}{d\,\mbox{ln} M} \, f(\sigma(M,z)) ~, 
\end{equation}
where the function $f (\sigma)$ represents the fraction of mass that has collapsed to form halos per unit interval in $\ln \sigma^{-1}$ and, if all the mass is assumed to be inside halos, it verifies $\int f(\sigma)  d\ln{\sigma^{-1}} = 1$.  Here, $\sigma(M,z)$ is
\begin{equation}
\sigma^{2}(M,z) = \left(\frac{D(z)}{D(0)}\right)^{2} \int \frac{dk}{k}\frac{k^{3} \, P(k)}{2\pi^{2}}
|W(kR)|^{2}\label{sigma} ~,
\end{equation} 
where $W(kR)$ is the window function and we consider a spherical top-hat window on scale $R^3=3 \, M/4\pi \rho_{\rm m,0}$; $P(k)$ is the matter power spectrum computed using the code provided with Ref.~\cite{Eisenstein:1997jh}; and $D(z)$ is the growth factor which we approximate as~\cite{Lahav:1991wc, Carroll:1991mt}
\begin{equation}
D(z) =
\left(\frac{(5/2)\Omega_{\rm m}(z)}{\Omega^{4/7}_{\rm m}(z) -
  \Omega_{\Lambda}(z) + (1+\Omega_{\rm m}(z)/2) (1 +
  \Omega_{\Lambda}(z)/70)}\right)\frac{1}{1+z}  ~,
\end{equation}   
with $\Omega_{\Lambda}(z)=1-\Omega_{\rm m}(z)$. 

The determination of the halo mass function from observational data is  challenging~\cite{Eke:2005za, Rines:2008tr, Vikhlinin:2008ym, Rozo:2009jj}.  Nevertheless it can be inferred from N-body simulations and analytical models.  The first analytical attempt was developed by Press and Schechter~\cite{Press:1973iz, Bond:1990iw}, who assumed that the fraction of mass in halos more massive than a given mass is related to the fraction of the volume in which the smoothed initial density field is above some threshold.  However, this model systematically underpredicts the number of halos at high masses and redshifts and overpredicts it for low masses at low redshifts~\cite{Springel:2005nw, Heitmann:2006hr, Lukic:2007fc, Watson:2012mt}.  Extending the spherical collapse model to an ellipsoidal collapse improved considerably the fits~\cite{Sheth:1999mn, Sheth:1999su, Sheth:2001dp}.  The mass function introduced by Ref.~\cite{Sheth:1999mn} (ST), using the GIF simulations~\cite{Kauffmann:1998gz}, is determined from the function    
\begin{equation}
\label{eq:ST}
f_{\rm ST}(\sigma) = A \, \sqrt{\frac{2 a}{\pi}} \, \left(1 + \left(\frac{a \, \delta_{\rm c}^2}{\sigma^2} \right)^{q} \right)\left(\frac{\delta_{\rm c}}{\sigma}\right) \, e^{-\frac{a \, \delta_{\rm c}^2}{2 \sigma^2}} ~,  
\end{equation}
with $a=0.75$, $q=0.3$ and $A=0.322$ and where $\delta_{\rm c} = 1.674$ is the critical overdensity required for collapse at $z=0$~\cite{Lilje:1991ym, White:1992ri, Kochanek:1994vw, Eke:1996ds}. In Ref.~\cite{Jenkins:2000bv}, it was suggested that the halo mass function could have an universal form in terms of the rms of matter fluctuations, $\sigma(M,z)$.  However, although the proposed parametrization leads to some improvements over that of ST, it cannot be extrapolated beyond the range where the fit was performed.  It was later shown that, although small, the departures from universality in redshift were not negligible~\cite{White:2002at}.  Nevertheless, the size of these deviations depend on the used halo finder, being larger for computations based on the SO algorithm, which tend to predict fewer halos~\cite{Tinker:2008ff}.

Several other fitting functions have been proposed using the results of numerical N-body simulations~\cite{Reed:2003sq, Warren:2005ey, Reed:2006rw, Lukic:2007fc, Cohn:2007xu, Tinker:2008ff, Crocce:2009mg, Courtin:2010gx, Bhattacharya:2010wy, Angulo:2012ep, Watson:2012mt}.  In this work, in addition to that of ST, we consider the parametrizations of Ref.~\cite{Tinker:2008ff}, with the results, based on the simulations with the SO algorithm, of Ref.~\cite{Tinker:2008ff} (T08) and Ref.~\cite{Watson:2012mt} (W13).

In the work of T08, the halo abundances are calibrated in the mass range $\sim 10^{11} - 10^{15} \, h^{-1} M_\odot$ up to $z \simeq 2$, and the function $f(\sigma)$, computed for overdensities in the range $200 \leq \Delta \leq 3200$, is parametrized as~\cite{Tinker:2008ff} 
\begin{equation}
\label{eq:fT08}
f_{\rm T08}(\sigma, z) = A(z) \, \left[\left(\frac{\sigma}{b(z)}\right)^{-a(z)} + 1 \right] \,
e^{-\gamma/\sigma^{2}}  ~,
\end{equation}
where the functions  $A(z), a(z), b(z),$ and $\gamma$ were parametrized as a simple scaling of the z=0 fitting parameters as~\cite{Tinker:2008ff},  
\begin{eqnarray}
\label{eq:T08param}
A(z) & = & A_{0}(1+z)^{-0.14} ~, \\
a(z) & = & a_{0}(1+z)^{-0.06} ~, \\
b(z) & = & b_{0}(1+z)^{-\alpha} ~, \\
\gamma & = & 1.19 ~, \\
\log_{10} \alpha(\Delta) & = & -\left[\frac{0.75}{\log_{10}(\Delta/75)}
  \right]^{1.2} ~,
\end{eqnarray}
where, for  $\Delta=200$, $A_{0}=0.186$, $a_{0}=1.47$ and $b_{0}=2.57$.

\begin{table}[t]
\begin{center}
\begin{tabular}{| l l l l l l l l | }
\hline
$f(\sigma, z)$ & $\Omega_{\rm m,0}$ & $\Omega_\Lambda$ & $\Omega_{\rm b,0}$
& $h$ & $n_s$ & $\sigma_{8}$ & $\Delta$ \\ \hline \hline
ST~\cite{Sheth:1999mn} & 0.3 & 0.7 & - & 0.7 & - & 0.9 & $\Delta_{\rm vir}$ \\ \hline  
T08~\cite{Tinker:2008ff} & 0.3 & 0.7 & 0.04 & 0.7 & 1.0 & 0.9 & 200 \\ \hline 
W13~\cite{Watson:2012mt} & 0.27 & 0.73 & 0.044 & 0.7 & 0.96 & 0.8 & 200 \\ \hline 
\end{tabular}
\end{center}
\caption{Parameters used in the simulations we consider to parametrize the halo mass function $f(\sigma,z)$, where $n_s$ is the scalar spectral index and $\sigma_8$ is the rms linear mass fluctuation within a sphere of radius $8~h^{-1}$Mpc extrapolated to $z = 0$.}
\label{tab:f}
\end{table}

\begin{figure}[t]
\includegraphics[width=\textwidth]{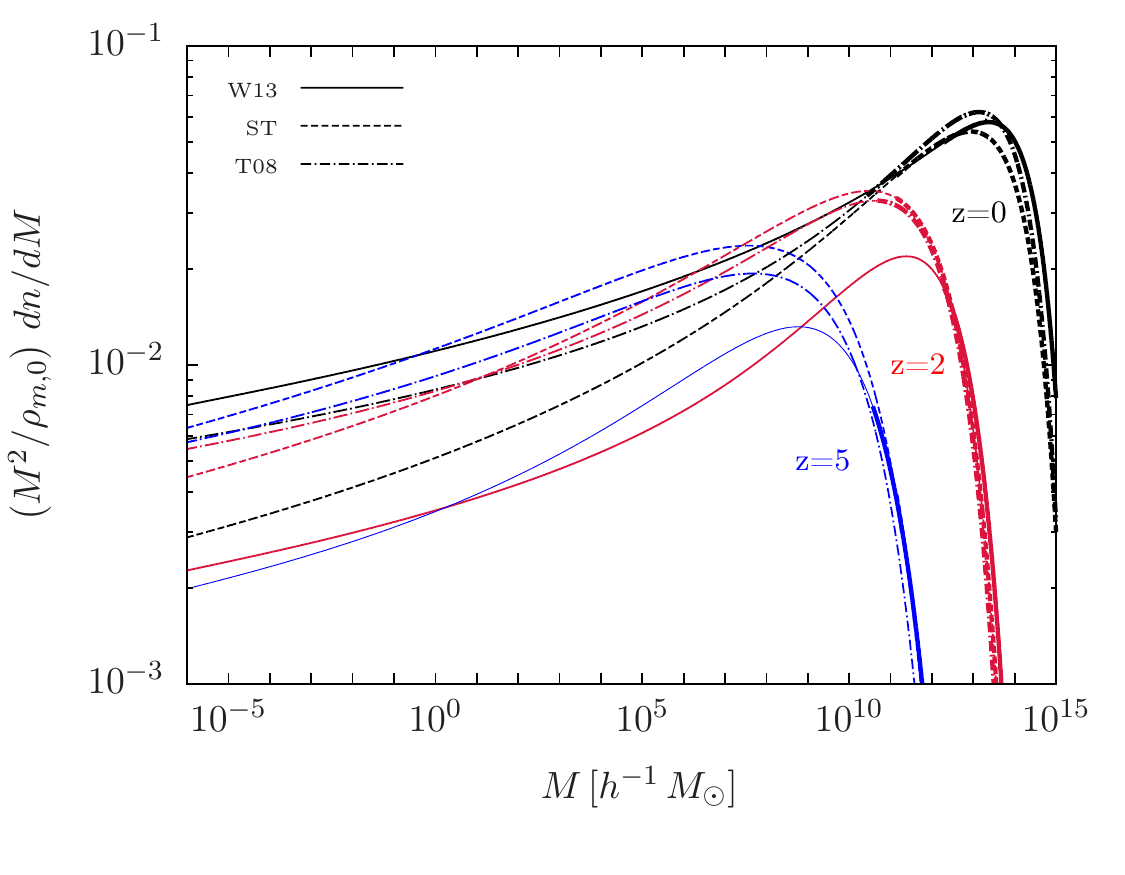}
\caption{The halo mass function at three redshifts, $z=0,2,5$ (black, red and blue lines, respectively), for the three simulations we consider: ST~\cite{Sheth:1999mn} (dashed lines), T08~\cite{Tinker:2008ff} (double-dot-dashed lines) and W13~\cite{Watson:2012mt} (solid lines). Thinner lines are used to indicate masses outside the range probed by the simulations.     
} 
\label{fig:dndM}
\end{figure}

On the other hand, the simulations in W13 allowed resolving CDM halos with masses $M \gtrsim 10^5 \, h^{-1} M_\odot$ up to redshifts $z\simeq 30$, and the parametrization for the mass function $f(\sigma,z)$ for a given $\Delta$ is given by~\cite{Watson:2012mt}  
\begin{equation}
\label{eq:fW13}
f_{{\rm W13}, \Delta}(\sigma,z) = \Gamma(\Delta;\sigma,z) \, f_{{\rm W13}, 178}(\sigma,z) ~, 
\end{equation}
where
\begin{equation}
\Gamma(\Delta; \sigma,z) = \mu^{d(z)} \, {\rm exp}[(\mu-1) \, (0.023 - 0.072/\sigma^{2.13})] ~,
\end{equation}
with $\mu=\Delta/178$, $d(z)=-0.456 \, \Omega_{\rm m}(z)-0.139$ and $f_{{\rm W13}, 178}(\sigma,z)$ is defined as in Eq.~(\ref{eq:fT08}), with the functions $A(z)$, $a(z)$, $b(z)$ and $\gamma$ parametrized for $\Delta=178$ as    
\begin{eqnarray}
A(z) & = & \Omega_{\rm m}(z) \, (1.097 (1 + z)^{-3.216} + 0.074) ~, \nonumber \\ 
a(z) & = & \Omega_{\rm m}(z) \, (5.907 (1 + z)^{-3.599} + 2.344) ~, \nonumber \\ 
b(z) & = & \Omega_{\rm m}(z) \, ( 3.136 (1 + z)^{-3.058} + 2.349) ~, \nonumber \\ 
\gamma & = & 1.318 ~.
\end{eqnarray}

In Fig.~\ref{fig:dndM} we show the parametrization of the halo mass function at three redshifts, $z=0,2,5$ for the three simulations we consider. All the cosmological parameters adopted are indicated in Tab.~\ref{tab:f}.

\subsection{Uncertainties in $\xi^{2}(z)$}
\label{subsec:uncertainties}

Once we have all ingredients to compute the neutrino flux, in this section we show the results for the enhancement factor for different combinations of parametrizations for $c_\Delta(M,z)$ and $f(M,z)$.  As we are using simulations which use either an overdensity $\Delta(z)=200$ or $\Delta(z)=\Delta_{\rm vir}(z)$, not all pairs of results can be combined in a consistent way.  Therefore, we consider four combinations for the concentration parameter and the halo mass function, which are indicated in Tab.~\ref{tab:combinations}.  

\begin{figure}[t]
	\hspace{-0.17cm}\includegraphics[width=0.67\textwidth]{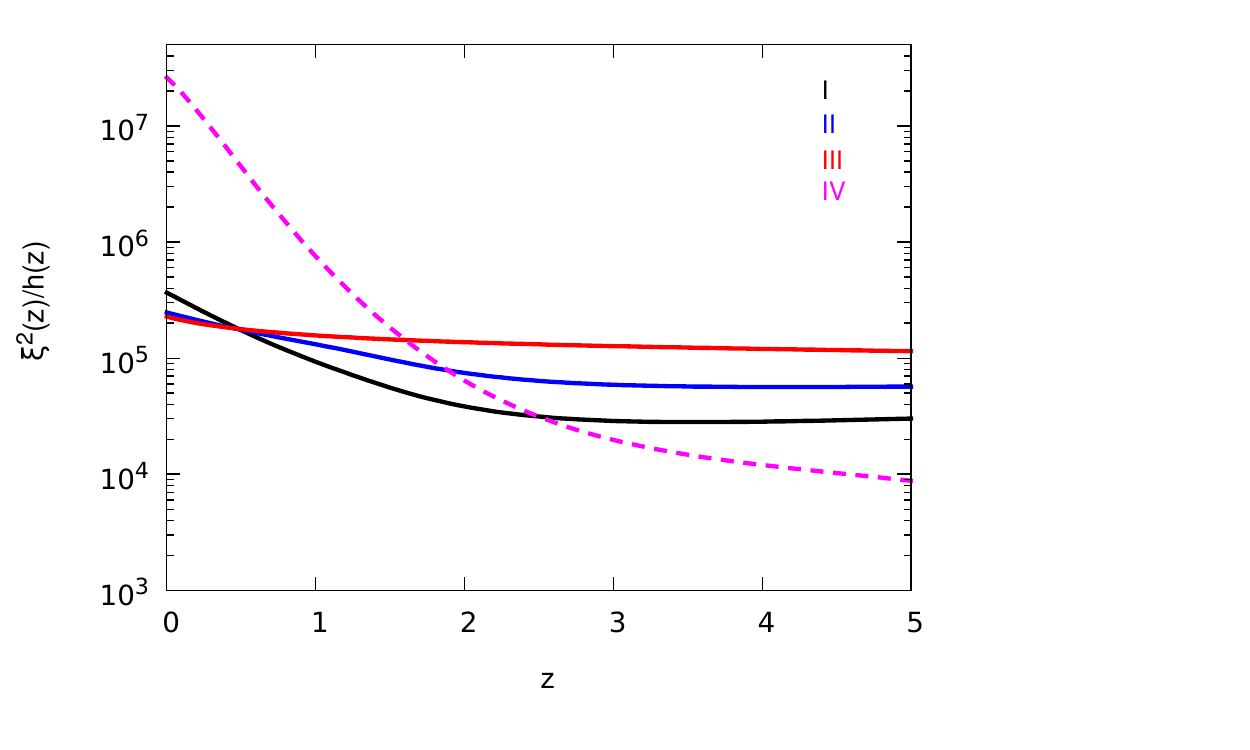}
	\hspace{-2.6cm} 
	\includegraphics[width=0.67\textwidth]{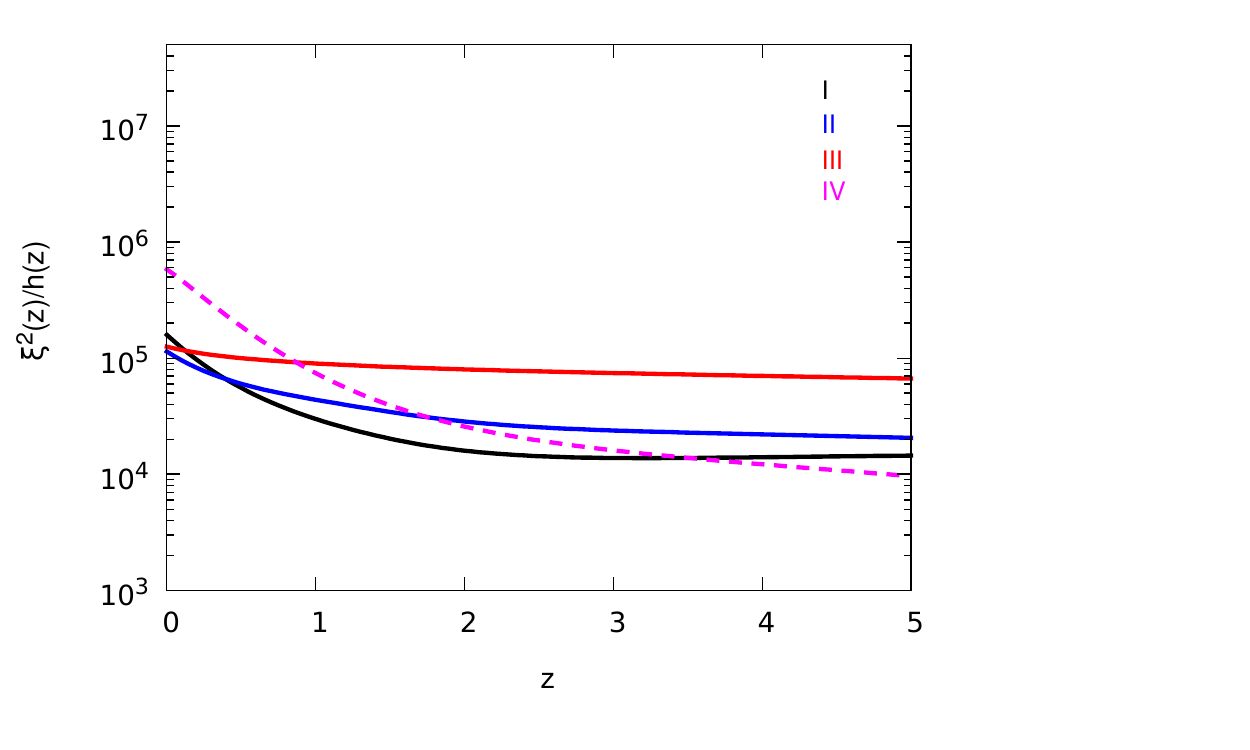}
	\caption{Left panel: The enhancement factor $\xi^2(z)/h(z)$ for all combinations described in Tab.~\ref{tab:combinations}: I (lower black solid line), II (middle blue solid line), III (upper red solid line), IV (magenta dashed line).  The results in the rest of the paper are shown for the combination I.  Right panel: Result of introducing a cutoff in $c_\Delta(M,z)$ at $M_{\rm cut}=10^5 \, M_\odot$ (see the text).}	
	\label{fig:xi2}
\end{figure}

In Fig.~\ref{fig:xi2}, we plot the enhancement factor $\xi^2(z)/h(z)$, Eq.~(\ref{eq:xi2}), as a function of redshift, for these four combinations (left panel) and in addition, we show (right panel) the result of modifying the parametrizations for the concentration parameter by introducing a cutoff at $M_{\rm cut}=10^5 \, M_\odot$, such that   
\begin{equation}
\label{eq:mcut}
c_\Delta(M,z) = c_\Delta(M_{\rm cut},z) \hspace{5mm} ; \hspace{5mm}
\forall M < M_{\rm cut} ~.
\end{equation}
Let us note, however, that in the case of P12, the concentration--mass relation exhibits a flattening towards low halo masses, so this cut has a smaller effect. 

As one can see from the left panel, the main differences are driven by differences in the parametrizations of the concentration parameter and how they extrapolate to low masses and high redshifts. The combination IV considers a parametrization for the concentration parameter (DM14) which represents a linear relation in the $\log_{10} c_\Delta - \log_{10} M$ plane and the concentration grows with mass at low redshifts.  On the contrary, P12 and B01 present a flattening at low masses (more pronounced for P12 due to its dependence on $\sigma(M,z)$, which also presents a flattening at low masses) and hence, the combinations I, II, III produce a similar redshift dependence, although for III,  $\xi^2(z)/h(z)$ is a factor of a few larger than for I and II at all redshifts. In part, this can be understood from the fact that $c_{\textrm{vir}} > c_{200}$. In the right panel, we also show the results when a cut in $c_\Delta(M,z)$ is introduced at $M_{\rm cut} = 10^5 \, M_\odot$, as described above.  In this case, the results with all combinations are much more similar (within an order of magnitude at all redshifts), which shows the impact of extrapolations well beyond the range probed by the simulations. As mentioned above, the effect of this cut at low redshifts in the case of P12 (and also of B01) is much smaller than for DM14, which results in making all results alike.  

Finally, let us note that in this paper we have not included the effect of substructure within halos.  As the DM annihilation signal is proportional to the square of the DM density, the existence of subhalos within larger halos may boost this signal. In fact, this boost factor is estimated to be larger for larger halos, where more substructure is expected~\cite{Taylor:2002zd, Strigari:2006rd, Pieri:2007ir, Lavalle:1900wn, Kamionkowski:2008vw, Kuhlen:2008aw, Springel:2008cc, Martinez:2009jh, Zavala:2009zr, Kamionkowski:2010mi, Charbonnier:2011ft, SanchezConde:2011ap, Pinzke:2011ek, Gao:2011rf, Fornasa:2012gu, Kuhlen:2012ft, Nezri:2012tu, Anderhalden:2012jc, Zavala:2013lia, Sanchez-Conde:2013yxa}. Nevertheless, for the cases we consider, the cosmological flux from DM annihilations is dominated by the more numerous low-mass halos (see also Ref.~\cite{Ullio:2002pj}) and hence, the impact on the final flux is expected to be moderate. Some studies find a boost factor of a few~\cite{Kamionkowski:2010mi, SanchezConde:2011ap, Fornasa:2012gu, Sanchez-Conde:2013yxa}, although some others find a boost factor an order of magnitude larger~\cite{Springel:2008cc, Pinzke:2011ek, Gao:2011rf}, which also illustrates the large uncertainties involved in obtaining these estimates. Bearing in mind this discussion, the relatively large uncertainties from the modeling of the smooth signal and the fact that it represents the minimum expected signal, we decide not to include the substructure boost.  In this regard, our results are conservative. 

In the rest of the paper, we use combination I as the representative of our results\footnote{Note that the parametrizations used in I depend on $\sigma(M,z)$ so, in principle, a cut in the concentration parameter for low masses is not fully justified. In any case, the differences are at the level of a factor of a few.}. Let us notice that the main contribution to the neutrino flux comes from low redshifts, so combinations I, II and III result in similar expectations. On the other hand the extrapolations implied by the combination IV represent a very optimistic case, which again we stress is highly disfavored by the recent findings for low-mass halos.

\begin{table}[t]
	\begin{center}
		\begin{tabular}{| l l l l | }
			\hline
			& $c_\Delta(M, z)$ & $f(\sigma,z)$ & $\Delta$ \\
			\hline \hline
			I & P12\cite{Prada:2011jf} & W13~\cite{Watson:2012mt} & 200 \\ \hline
			II & P12~\cite{Prada:2011jf} & T08~\cite{Tinker:2008ff} & 200 \\ \hline
			III & B01~\cite{Bullock:1999he, Dutton:2014xda} & ST~\cite{Sheth:1999mn} & $\Delta_{\rm vir}$ \\ \hline
			IV & DM14~\cite{Dutton:2014xda} & W13~\cite{Watson:2012mt} & 200 \\ \hline 
		\end{tabular}
	\end{center}
	\caption{Combinations of parametrizations for $c_\Delta(M,z)$ and $f(\sigma,z)$ considered in this work in order to estimate the enhancement factor $\xi^2(z)$.} 
	\label{tab:combinations}
\end{table}

\subsection{The Milky Way contribution}

\begin{figure}[t]
\includegraphics[width=\textwidth]{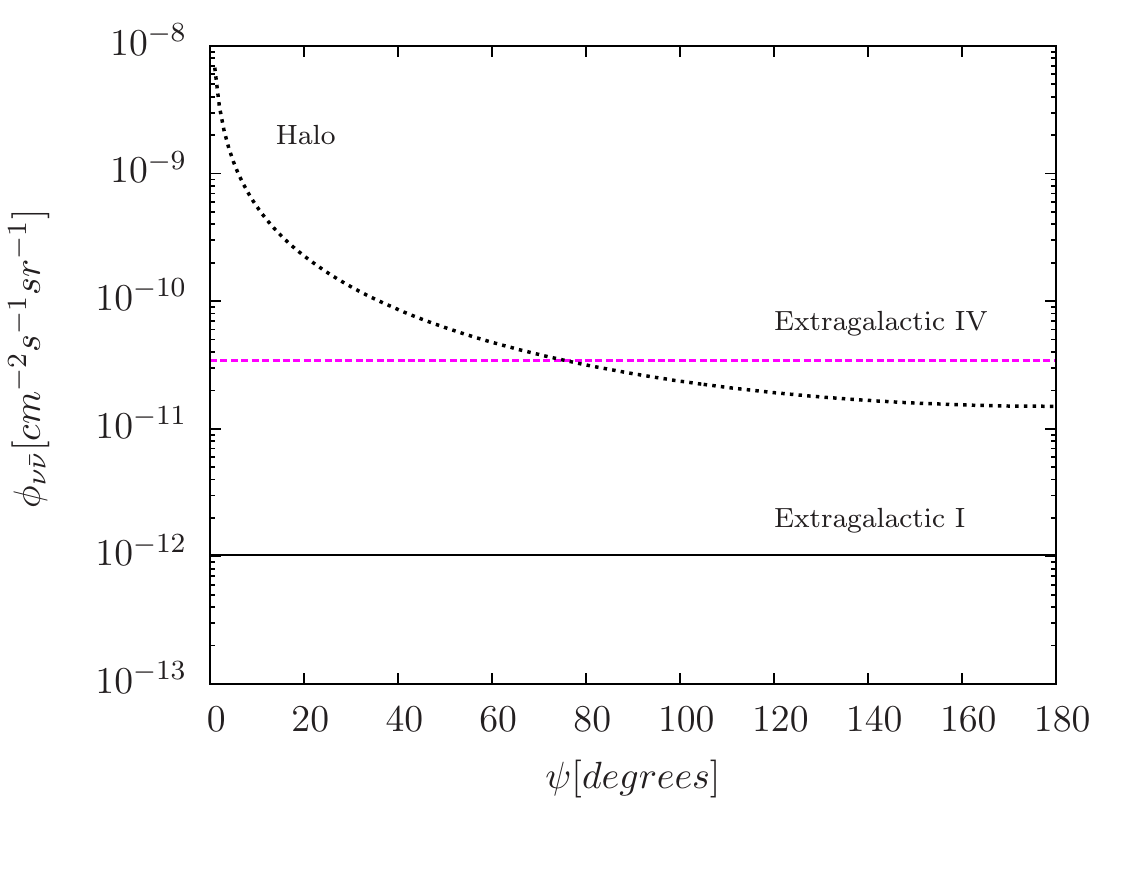}
\caption{Averaged neutrino flux from DM annihilations in a solid angle $\Delta\Omega = 2\pi \, (1-\cos\psi)$ around the galactic center and in the logarithmic energy bin $\Delta\log_{10} E_0 = \log_{10} \left(m_{\rm DM}/E_0\right) = 0.3$, as a function of the angular distance from the galactic center, assuming the final state $\nu\bar \nu$, $m_{\rm DM}=1$~TeV and $\langle \sigma v\rangle= 3\times 10^{-26}\,{\rm cm}^3\,{\rm s}^{-1}$.  The black dotted line represents the contribution from DM annihilations in the Milky Way halo, while the solid black (dashed magenta) line, refers to cosmological DM annihilations for the combination I (IV) to determine the enhancement factor.   
}
\label{fig:cosmo_vs_GC}
\end{figure}

In this paper we consider the cosmological signal to be the only contribution to the neutrino flux from DM annihilations. Nevertheless, there exists another contribution, potentially dominating the cosmological signal, stemming from DM annihilations in the halo of our own galaxy.  We consider the Milky Way DM halo to be described by the NFW profile, Eq.~(\ref{eq:nfw1}), with $r_s=20$~kpc and with a local DM density $\rho(R_\odot)=0.4\,{\rm GeV}\,{\rm cm}^{-3}$, where $R_\odot=8.5$~kpc is the distance from the Sun to the galactic center.  

The differential energy flux, neglecting neutrino flavor mixing, in the direction forming an angle $\psi$ with respect to the galactic center is given by
\begin{equation}
\frac{d \phi_{\nu_\beta}(E_0)}{dE_{0}} \Big|_{{\rm GC}, \psi} =
\frac{1}{4\pi } \frac{\langle\sigma \upsilon\rangle}{2 \, m^2_{\rm DM}} \int_0^\infty  \rho^2[r(\ell,\psi)] \,d\ell \, \sum_{i} {\rm Br}_{i} \frac{dN_{\nu_\beta, i}(E_0)}{dE_0}   ~,
\end{equation}
where $r(\ell,\psi)=\sqrt{R_\odot^2-2\ell R_\odot\cos\psi+\ell^2}$. The effect of neutrino oscillations can be included using Eq.~(\ref{eq:prob}).  

We compute the average flux in a solid angle $\Delta\Omega = 2\pi \, (1-\cos\psi)$ and in the logarithmic energy bin $\Delta\log_{10} E_0 = \log_{10} \left(m_{\rm DM}/E_0\right) = 0.3$, and in Fig.~\ref{fig:cosmo_vs_GC} we compare the angular dependence with respect to the direction of the Milky Way center of the galactic and the extragalactic neutrino signals (assuming DM annihilations into $\nu\bar\nu$) for our two extreme combinations for the enhancement factor, I and IV.  In order to calculate the fluxes, we assume $\langle \sigma v \rangle =3\times 10^{-26}\,{\rm cm}^3 \, {\rm s}^{-1}$ and $m_{\rm DM}=1$ TeV, although the relative size of the halo and the extragalactic components depends little on this choice.  

As apparent from the plot, close to the galactic center the total neutrino flux is dominated by the halo component.  Nevertheless, at large angular distances the  extragalactic component amounts to at least $\sim$10\% of the total flux.  Furthermore, depending on the structural properties of the individual halos and on the abundance of halos as a function of mass and redshift, the extragalactic component can provide a contribution comparable to the flux at large angular distances from the galactic center.  This comparison has also been recently discussed in the context of gamma-rays~\cite{Fornasa:2012gu, Ng:2013xha, Ackermann:2015tah}.

\section{Event spectra in neutrino telescopes}
\label{sec:event}

In neutrino telescopes, neutrinos are detected via neutrino-nucleon interactions which take place either inside or outside the detector. In the case of muon neutrinos (and antineutrinos), they can interact via a charged-current (CC) interaction and produce a muon outside the instrumented detector, which could enter the detector and be detected as a through-going track event.  On the other hand, showers are produced via CC interactions of electron or tau\footnote{In our simplified analysis, we neglect the fact that a tau decaying into a muon and a neutrino-antineutrino pair, which has a branching fraction of 17.41\%, produces a track.} neutrinos (and antineutrinos) and via neutral-current (NC) interactions of all neutrino flavors.  In this work, we consider these two types of events in km$^3$ \v{C}erenkov neutrino telescopes such as IceCube, ANTARES or the future KM3NeT and we evaluate the reach of these detectors to set limits on the DM annihilation cross section using the cosmic signal after 10~years of data-taking.

\subsection{Through-going muons}
\label{subsec:tracks}

When neutrinos interact via CC interactions with the nucleons in the surrounding material of the detector, the generated muons can travel up to the detector, which enhances the effective detector area for high-energy muon neutrinos.  In their passage through the material, such muons lose energy before they reach the detector.  The muon average energy loss over a distance $dx$ during their travel through a medium with density $\rho$, is given by 
\begin{equation}
\label{eq:dedx}
-\frac{dE_{\mu}}{dx}  = \rho \, \left[\alpha(E_{\mu}) + \beta(E_{\mu}) E_{\mu}\right] ~, 
\end{equation}
where $\alpha(E_{\mu})$ describes the ionization energy loss and $\beta (E_{\mu})$ accounts the energy loss due to bremsstrahlung, $e^{+}e^{-}$ pair production and photonuclear interactions~\cite{Groom:2001kq}.  These quantities vary very smoothly with energy and we approximate them as constants, $\alpha \simeq 2 \times  10^{-3} \, {\rm GeV cm}^{2}/{\rm g}$ and $\beta \simeq 3 \times 10^{-6} \, {\rm cm}^{2}/{\rm g}$~\cite{Erkoca:2009by}.  

The initial muon energy at its production point, $E^{i}_{\mu}$, and the final energy, $E_{\mu}$, of a muon when it is detected after traveling a distance $x$ in a surrounding material of the detector,
are related as 
\begin{equation}
\label{eq:emui}
E^{i}_{\mu} = e^{\beta \rho x} E_{\mu}+(e^{\beta \rho x}-1)\frac{\alpha}{\beta} ~,  
\end{equation}
and the average range after which the muon energy drops below a detector threshold energy $E_{\rm th}$ is given by\footnote{In this formula we have assumed continuous energy loses.  However, above a few TeV ($E_\mu > \alpha/\beta$) stochastic energy losses start being important and thus, Eq.~(\ref{eq:range-muon}) overestimates the muon range for large energies.}
\begin{equation}
\label{eq:range-muon}
R_{\mu}(E^{i}_{\mu},E_{\rm th}) = \frac{1}{\beta \rho} \mbox{ln}\left(\frac{\alpha + \beta \, E^{i}_{\mu}}{\alpha + \beta \,  E_{\rm th}} \right) ~.
\end{equation}
In addition, taking muon decay into account, the probability for a muon to arrive at the detector is given by~\cite{Erkoca:2009by}
\begin{equation}
P(E^{i}_{\mu},E_{\mu}) = \left(\frac{E_{\mu}}{E^{i}_{\mu}}\right)^{\Gamma} \left(\frac{\alpha + \beta E^{i}_{\mu}}{\alpha + \beta E_{\mu}} \right)^{\Gamma} ~, 
\end{equation}
where $\Gamma = m_{\mu}/(\tau_\mu \alpha \rho)$, and $m_\mu$ and $\tau_\mu$ are the muon mass and lifetime, respectively.

The induced flux of through-going muon events can be obtained by convoluting the incoming neutrino flux with the differential neutrino cross section and taking into account the shift to lower energies due to energy losses during the propagation in the surroundings of the detector\footnote{Here, we have not included the attenuation of neutrinos in their passage through the Earth, which at the energies relevant to this analysis is negligible.}.  The result is~\cite{Erkoca:2009by, Covi:2009xn}   
\begin{eqnarray}
\label{eq:muflux}
\frac{d\phi_{\mu}}{dE_{\mu}} & = & \int d\Omega \int^{m_{\chi}}_{E_{\mu}}
dE_{\nu_{\mu}} \int_{0}^{R_\mu (E_{\mu}^i,E_\mu)} dr \, e^{\beta \rho r} \, \frac{d\phi_{\nu_{\mu}}(E_{\nu_{\mu}})}{dE_{\nu_{\mu}}}
 \left[ n_{\rm p} \frac{d\sigma^{\nu \rm p}_{\rm CC}(E_{\nu_{\mu}},E^{i}_{\mu})}{dE_{\mu}} +   
    ({\rm p} \rightarrow {\rm n}) \right] \, P(E^{i}_{\mu},E_{\mu}) \nonumber \\  
 & & + (\nu \rightarrow \bar{\nu}) ~,
\end{eqnarray}
where $E_\mu^i$ is given in Eq.~(\ref{eq:emui}), while  $n_{\rm p} = \rho N_{A}\langle Z/A \rangle$ and $n_{n}$=$\rho N_{A}(1-\langle Z/A \rangle)$ are the number density of protons and neutrons\footnote{Note that $+ ({\rm p} \rightarrow {\rm n})$ indicates that one should add the corresponding contribution from neutrons.}, respectively, with $N_{A}$ the Avogadro number, $\rho$ the density of the material and $\langle Z/A \rangle$ the average ratio of the proton number and the mass number, which we take $\langle Z/A \rangle= 8/18$ as is the case of water or ice.  Besides, the CC deep inelastic differential cross section, $d\sigma^{\rm p,n}_{\nu}/dE_\mu$ is approximately given by~\cite{Strumia:2006db, Barger:2007xf}
\begin{equation}
\label{eq:dsigma}
\frac{d\sigma^{\nu \rm p,n}_{\rm CC} (E_{\nu_\mu}, E_\mu)}{dE_\mu} =
\frac{2 \, m_{\rm p,n} \, G^{2}_{\rm F}}{\pi}\left(a_{\rm CC}^{\nu \rm
  p,n}+b_{\rm CC}^{\rm p,n} \frac{E_\mu^2}{E_\nu^2} \right) ~,
\end{equation}
with $a_{\rm CC}^{\nu \rm p,n}=0.15, 0.25$,  $b_{\rm CC}^{\nu \rm p,n}=0.04, 0.06$ and $a_{\rm CC}^{\bar{\nu} \rm p,n}=b_{\rm CC}^{\nu \rm p,n}$, $b_{\rm CC}^{\bar{\nu} \rm p,n}=a_{\rm CC}^{\nu \rm p,n}$, and $G_{\rm F}$ is the Fermi constant. 

The differential induced-muon rate in the detector, $dN_\mu/dE_\mu dt$, is then calculated multiplying the induced flux of through-going muons, Eq.~(\ref{eq:muflux}), by the effective area $A_\mu(E_\mu)$, which is defined as the ratio of the rate of reconstructed events and the muon flux and includes the geometry of the detector and the detection efficiency.  This area is similar to the geometrical area, although it has some energy dependence.  For simplicity,  we assume in this work an effective area $A_\mu(E_\mu) = 1 \, {\rm km}^2$.  Furthermore, in order to avoid the huge atmospheric muon background, we do not include downgoing contained muon events and only take into account the events produced by upward-going neutrinos, as described above. Therefore, we integrate in a solid angle of $2\pi$.

\begin{figure}[t]
\includegraphics[width=\textwidth]{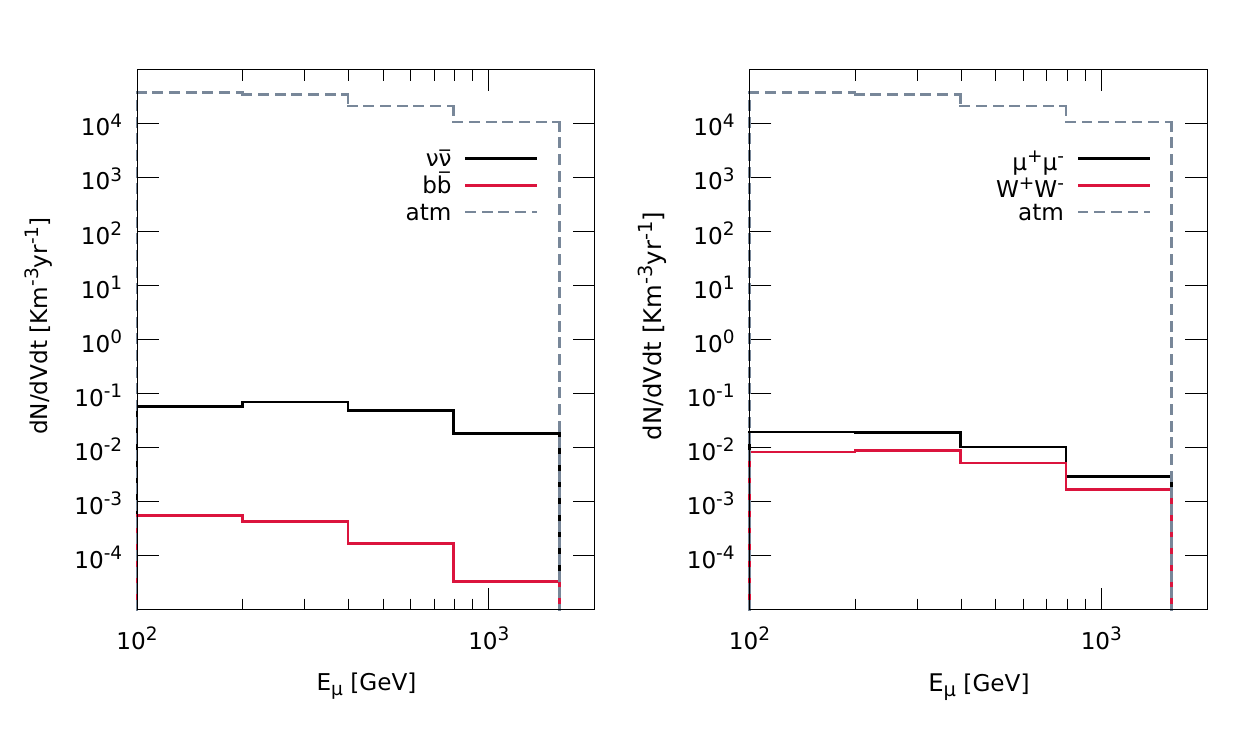}
\caption{The event spectra for through-going events for $m_{\rm DM}=1~{\rm TeV}$ and $\langle\sigma \upsilon\rangle = 3 \times 10^{-26} \, {\rm cm}^3/{\rm s}$, for four annihilation channels and the combination I for the enhancement factor.  Left panel: Annihilations into $\nu\bar{\nu}$ (black upper line) and into $b\bar{b}$ (red lower line).  Right panel: Annihilations into $\mu^+\mu^-$ (black upper line) and into $W^+W^-$ (red lower line). We also show the expected background spectrum due to atmospheric neutrinos (dotted lines), using the conventional atmospheric neutrino fluxes from Ref.~\cite{Honda:2006qj} and the tau neutrino flux from charmed meson decays~\cite{Pasquali:1998xf}.  Note  that, because of the choice of the bin size, the last bin goes over 1~TeV.
}    
\label{fig:tgspectra}
\end{figure}

Lastly, we include the energy resolution for tracks,
\begin{equation}
\label{eq:ResEmu}
R_\mu (\log_{10} E_\mu, \log_{10} E_\mu^{\rm mes}) = \frac{1}{\sigma_{\log_{10} E_\mu} \, \sqrt{2 \pi}} \, \, e^{-\frac{(\log_{10} E_\mu - \log_{10} E_\mu^{\rm mes})^2}{2 \, \sigma_{\log_{10} E_\mu}^2}} ~,  
\end{equation}
where $E_\mu$ and $E_\mu^{\rm mes}$ are the true and the measured energies, respectively, and $\sigma_{\log_{10} E_\mu} = 0.3$~\cite{Resconi:2008fe}.  Finally, the measured differential muon rate reads:  
\begin{equation}
\label{eq:specEmu}
\frac{dN_\mu}{d\log_{10} E_\mu^{\rm mes} \, dt} = \int_{\log_{10} E_{\rm th}}^{\log_{10} m_{\rm DM}} d\log_{10} E_\mu \, R_\mu (\log_{10} E_\mu, \log_{10} E_\mu^{\rm mes}) \,
\frac{dN_\mu}{d\log_{10} E_\mu \, dt} ~, 
\end{equation}
where, in this simplified analysis, we have taken an energy threshold of 100~GeV. 

The predicted through-going event spectra for the exemplary annihilation final states $W^+W^-$, $\mu^+\mu^-$, $b\bar{b}$ and $\nu \bar{\nu} =(\nu_e \bar{\nu}_e + \nu_\mu \bar{\nu}_\mu + \nu_\tau \bar{\nu}_\tau)/3$  are shown in Fig.~\ref{fig:tgspectra}, for $m_{\rm DM}=1$~TeV and assuming the thermal value for the annihilation cross section, $\langle\sigma \upsilon\rangle = 3 \times 10^{-26} \, {\rm cm}^3/{\rm s}$, along with the expected background spectra due to the atmospheric neutrino flux.  In the plots we have taken logarithmic bins of equal size to the $1\sigma$ energy resolution.

\subsection{Showers}
\label{subsec:showers}

Neutrino-nucleon interactions occurring inside the detector also produce shower-like events.  The CC deep inelastic interaction of an electron neutrino (or antineutrino) off a nucleon produces a recoiling hadronic shower along with an electromagnetic shower caused by the propagation of the outgoing electron (or positron).  On the other hand, the CC deep inelastic interaction of a tau neutrino (or antineutrino) off a nucleon similarly produces a hadronic shower and accompanying tau, which at the energies of interest for this analysis, promptly decays in the detector and produces another shower\footnote{We neglect the the fact that a small fraction of taus, 17.41\%, decays into a muon and produces a track-like event.} (electromagnetic or hadronic, depending on the tau decay channel). Therefore, we assume that both, CC interactions of electron and tau neutrinos, deposit all the neutrino energy in the detector, which would allow a better reconstruction of the initial neutrino spectrum. Finally, showers are also produced via NC deep inelastic interactions of all neutrino (and antineutrino) flavors, although in this case, the outgoing neutrino is not detected and thus, only the recoiling hadronic shower is observed.  Therefore, the total rate of shower events per unit detector volume is given by~\cite{Covi:2009xn} 
\begin{eqnarray}
\label{eq:showers}
\frac{dN_{\rm sh}}{dE_{\rm sh} \, dt} & = & V_{\rm eff} \, \int d\Omega \left[\sum_{l=e,\tau}
  \left(\frac{d\phi_{\nu_l}(E_{\nu_l})}{dE_{\nu_l}} \times \left[ n_{\rm p} \sigma^{\nu \rm p}_{\rm CC}(E_{\nu_l}) + ({\rm p} \rightarrow {\rm n})\right]\right)_{E_{\nu_l}=E_{\rm sh}}\right. \nonumber \\   
 & & \left. + \sum_{l=e,\mu,\tau} \,\int^{m_{\chi}}_{E_{\rm sh}} dE_{\nu_{l}}\frac{d \phi_{\nu_{l}}(E_{\nu_{l}})}{dE_{\nu_{l}}} \times\left[ n_{\rm p} \frac{d\sigma^{\nu \rm p}_{\rm NC}(E_{\nu_{l}},E_{\rm sh})}{dE_{\rm sh}}+({\rm p} \rightarrow {\rm n})\right] \right] \nonumber\\ 
 & & +(\nu \rightarrow \bar{\nu}) ~,
\end{eqnarray}
where  the total cross-section for CC/NC interactions is approximately given by~\cite{Strumia:2006db, Barger:2007xf} 
\begin{equation}
\sigma^{\nu \rm p,n}_{\rm CC/NC}(E_{\nu_{l}}) =
\frac{2 \, m_{\rm p,n} \, G^{2}_{F}}{\pi}E_{\nu_{l}}\left(a_{\rm CC/NC}^{\nu \rm p,n}+\frac{1}{3} \,  b_{\rm CC/NC}^{\nu \rm p,n}\right) ~,
\end{equation}
with $a_{\rm NC}^{\nu \rm p,n}=0.058, 0.064$,  $b_{\rm NC}^{\nu \rm p,n}=0.022, 0.019$ and $a_{\rm NC}^{\bar{\nu} \rm p,n}=b_{\nu}^{\rm p,n}$, $b_{\rm NC}^{\bar{\nu} \rm p,n}=a_{\nu}^{\rm p,n}$.  The differential cross section $d\sigma^{\nu \rm p,n}_{\rm NC}(E_{\nu_{l}},E_{\rm sh})/dE_{\rm sh}$ is defined analogously to Eq.~(\ref{eq:dsigma}), with $E_{\rm sh} = E_{\nu_{l}} - E_l$.  Shower events are all contained events.  Thus, in this analysis we consider the effective volume to be approximately equal to the geometrical volume of the detector, $V_{\rm eff} = 1~{\rm km}^3$. 

Similarly to the through-going events, $dN_{\rm sh}/dE_{\rm sh} dt$ is the differential rate in terms of the true energy, $E_{\rm sh}$. To calculate the differential rate in terms of the measured energy, $E_{\rm   sh}^{\rm mes}$, we include the energy resolution for showers, 
\begin{equation}
\label{eq:ResEsh}
R_{\rm sh} (\log_{10} E_{\rm sh}, \log_{10} E_{\rm sh}^{\rm mes}) =
\frac{1}{\sigma_{\log_{10} E_{\rm sh}} \, \sqrt{2 \pi}} \, \, 
e^{-\frac{(\log_{10} E_{\rm sh} - \log_{10} E_{\rm sh}^{\rm mes})^2}{2 \, \sigma_{\log_{10} E_{\rm sh}}^2}} ~, 
\end{equation}
with $\sigma_{\log_{10} E_{\rm sh}} = 0.18$~\cite{Resconi:2008fe}. Therefore, the measured differential shower rate is given by  
\begin{equation}
\label{eq:specEsh}
\frac{dN_{\rm sh}}{d\log_{10} E_{\rm sh}^{\rm mes} \, dt} = \int_{\log_{10}
  E_{\rm th}}^{\log_{10} m_{\rm DM}} d\log_{10} E_{\rm sh} \, R_{\rm sh} (\log_{10} E_{\rm sh}, \log_{10} E_{\rm sh}^{\rm mes}) \, \frac{dN_{\rm sh}}{d\log_{10} E_{\rm sh} \, dt} ~,
\end{equation}
where we assume a shower energy threshold $E_{\rm th} = 100$~GeV.

The predicted shower event spectra for the four channels we consider in this paper are shown in Fig.~\ref{fig:tgspectra}, for $m_{\rm DM}=1$~TeV and assuming the thermal value for the annihilation cross section, $\langle\sigma \upsilon\rangle = 3 \times 10^{-26} \, {\rm cm}^3/{\rm s}$, along with the expected background spectra.  We take logarithmic bins of equal size to the $1\sigma$ energy resolution.

\begin{figure}[t]
	\includegraphics[width=\textwidth]{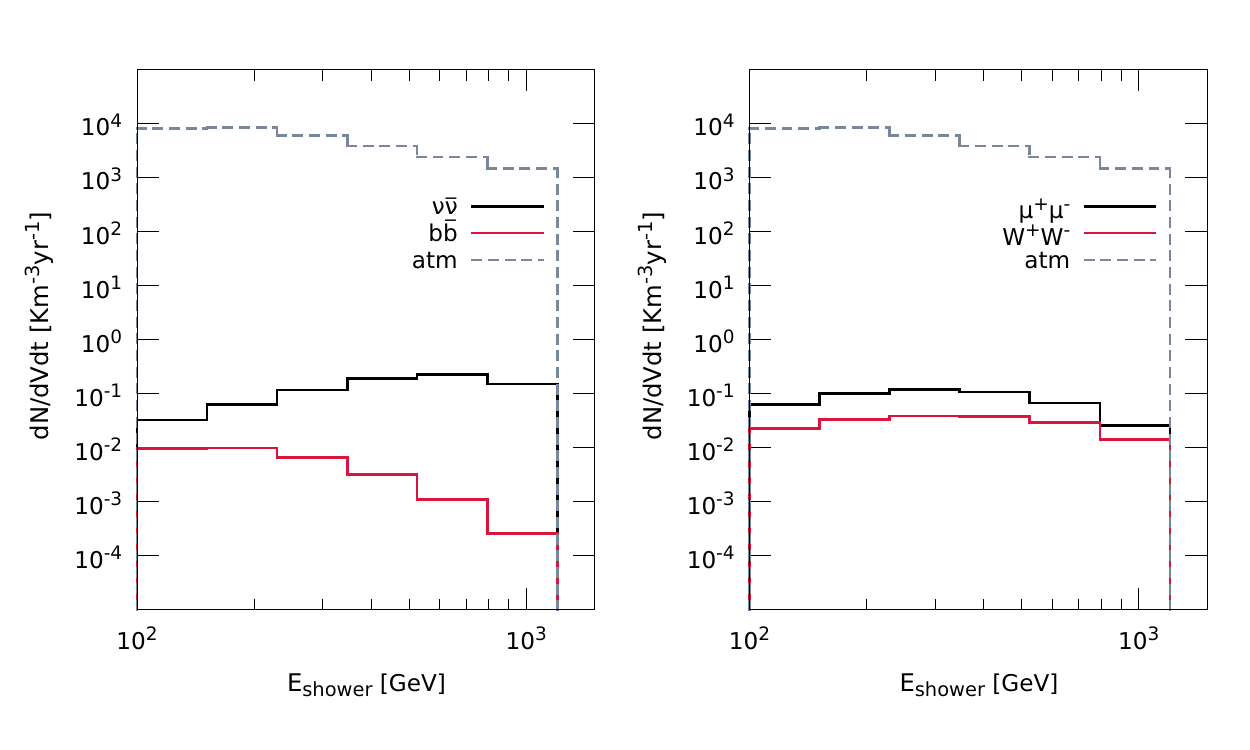}
	\caption{The event spectra for shower events for $m_{\rm DM}=1~{\rm TeV}$ and $\langle\sigma \upsilon\rangle = 3 \times 10^{-26} \, {\rm cm}^3/{\rm s}$, for four annihilation channels and the combination I for the enhancement factor.  Left panel: Annihilations into $\nu\bar{\nu}$ (black upper line) and $b\bar{b}$ (red lower line).  Right panel: Annihilations into into $\mu^+\mu^-$ (black upper line) and into $W^+W^-$ (red lower line).  We also show the expected background spectrum due to atmospheric neutrinos (dotted lines), using the conventional atmospheric neutrino fluxes from Ref.~\cite{Honda:2006qj} and the tau neutrino flux from charmed meson decays~\cite{Pasquali:1998xf}.  Note  that, because of the choice of the bin size, the last bin goes over 1~TeV.
	}    
	\label{fig:shspectra}
\end{figure}

\section{Future sensitivity to the dark matter annihilation cross section} 
\label{sec:bounds}

Here we present two simplified statistical analyses to forecast the future reach of neutrino telescopes to set limits on the cosmic DM annihilations.  We present results for 10~years of data-taking, for four exemplary channels, $W^+W^-$, $\mu^+\mu^-$, $b\bar{b}$ and $\nu \bar{\nu} = (\nu_e \bar{\nu}_e + \nu_\mu \bar{\nu}_\mu + \nu_\tau \bar{\nu}_\tau)/3$, and for the selected combination I for the enhancement factor $\xi^2(z)$, as discussed above. 

In order to set these limits, the potential signal has to be compared with the expected background.  At GeV-TeV energies, the main source of background, for the type of events discussed in this work, is the conventional flux of atmospheric neutrinos, which is originated after the interactions of cosmic-rays with the nuclei of the atmosphere and the subsequent decay of the produced pions and kaons.  Indeed, at energies above $\sim$100~GeV, the atmospheric neutrino flux is dominated by kaon decays, even though the number of kaons is smaller than that of pions at production.  This can be understood from the fact that most of the energy of the pions goes to the muon, whereas in the case of kaons it is approximately equally split between the daughter muon and neutrino.  Let us note that, although at energies below a few GeV the na\"{\i}ve ratio at production $\nu_\mu/\nu_e \simeq 2$ holds, at higher energies not all muons decay before reaching the ground and hence, that flavor ratio grows with energy. This has important consequences: the expected background for showers is smaller than that for through-going muon tracks, so the signal-to-noise ratio is larger for showers, as has already been pointed out~\cite{Mandal:2009yk, Lee:2012pz} and we will see next.  Finally, whereas electron neutrinos have no time to oscillate from the production point in the atmosphere to the detector, oscillations of atmospheric muon neutrinos into tau neutrinos do take place, in particular at energies below a few hundreds of GeV.  Thus, we take them into account as done in Ref.~\cite{Covi:2009xn}.  In this work, we take the conventional atmospheric neutrino fluxes from Ref.~\cite{Honda:2006qj} and include the contribution from charmed meson decays to the tau neutrino flux, which is the dominant one, using the parametrization of Ref.~\cite{Pasquali:1998xf}.

First we consider an aggressive analysis, for which we only assume statistical errors.  Therefore, we define a $\chi^2$ as
\begin{equation}
\chi^{2} = \sum_{i} \left(\frac{S_{i}(m_{\rm DM}, \langle\sigma \upsilon\rangle)}{\sqrt{B_{i}}}\right)^{2} ~,   
\end{equation}
where $S_i(m_{\rm DM}, \langle\sigma \upsilon\rangle)$ and $B_i$ are the signal and background events in the energy bin $i$.  

Then, we also include systematic errors on the atmospheric neutrino-induced events and perform a conservative analysis, following that in Ref.~\cite{FermiAbdo:2010dk}.  We restrict the DM signal to be smaller than the expected background (within errors) in all energy bins.  The limit on the annihilation cross section $\langle\sigma \upsilon\rangle$ is obtained by finding the bin $i$ with the largest signal/background ratio and equating it to 1, 
\begin{equation}
\left(\frac{S_i(m_{\rm DM}, \langle\sigma \upsilon\rangle)}{B_i + n (\sqrt{B_i} + \varepsilon B_i)}\right)_{\rm max} = 1 ~, 
\end{equation}
where we assume a 30\% systematic error on the normalization of the atmospheric neutrino flux, $\varepsilon = 0.3$, and $n=1.28$ correspond to a 90\% confidence level (C.L.) for an one-sided limit.

\begin{figure}[t]
\includegraphics[width=\textwidth]{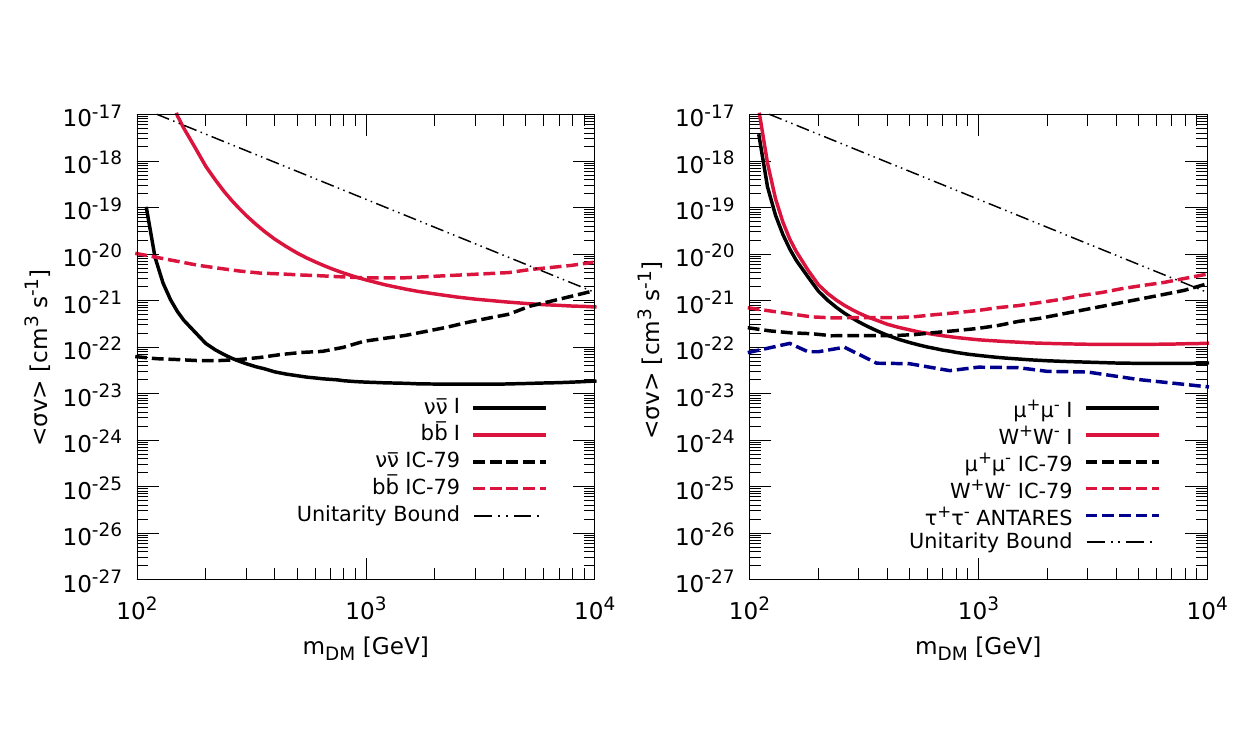} 
\vspace{-1.9cm} \\
\includegraphics[width=\textwidth]{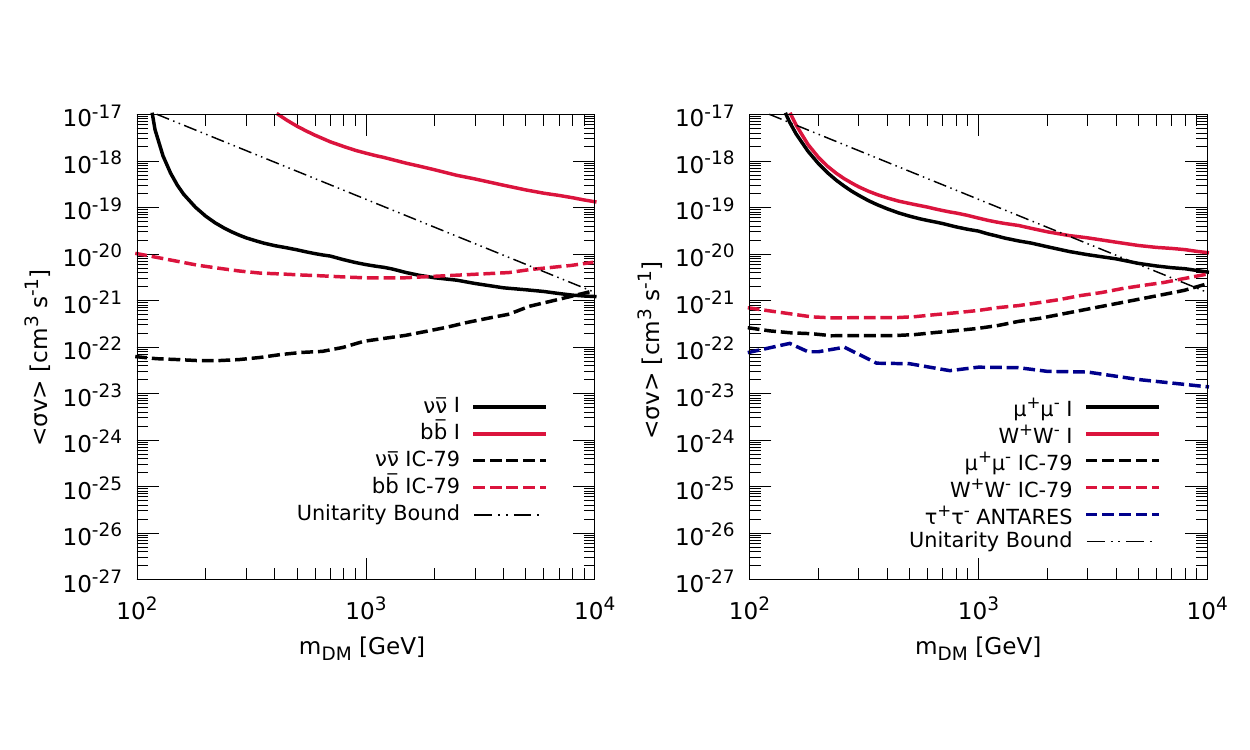}
\vspace{-1cm}
\caption{Aggressive (upper panels) and conservative (lower panels) limits on the DM annihilation cross section at 90\%~C.L. with through-going events after 10~years of data taking in a $1~{\rm km}^3$ neutrino telescope, for the combination I for the enhancement factor (solid lines). Left panels: Annihilations into $\nu\bar{\nu}$ (black solid lines) and into $b\bar{b}$ (red solid lines).  Right panels: Annihilations into $\mu^+\mu^-$ (black solid lines) and into $W^+W^-$ (red solid lines).  We also show the IC-79 90\% C.L. limits after 320 days of data (black and red dotted lines)~\cite{Aartsen:2013mla} from the galactic center and, on the right panels, shown as well are the ANTARES 90\% C.L. limits for DM annihilations into $\tau^+\tau^-$ from the Milky Way center after about 1300 days of data (blue dotted lines)~\cite{Hernandez-Rey:2014ssa, Zornoza:2014cra}. The unitarity limit~\cite{Griest:1989wd, Hui:2001wy} is shown as a double-dot-dashed line.     
}   
\vspace{5mm}
\label{fig:tglimits}
\end{figure}

\begin{figure}[t]
\includegraphics[width=\textwidth]{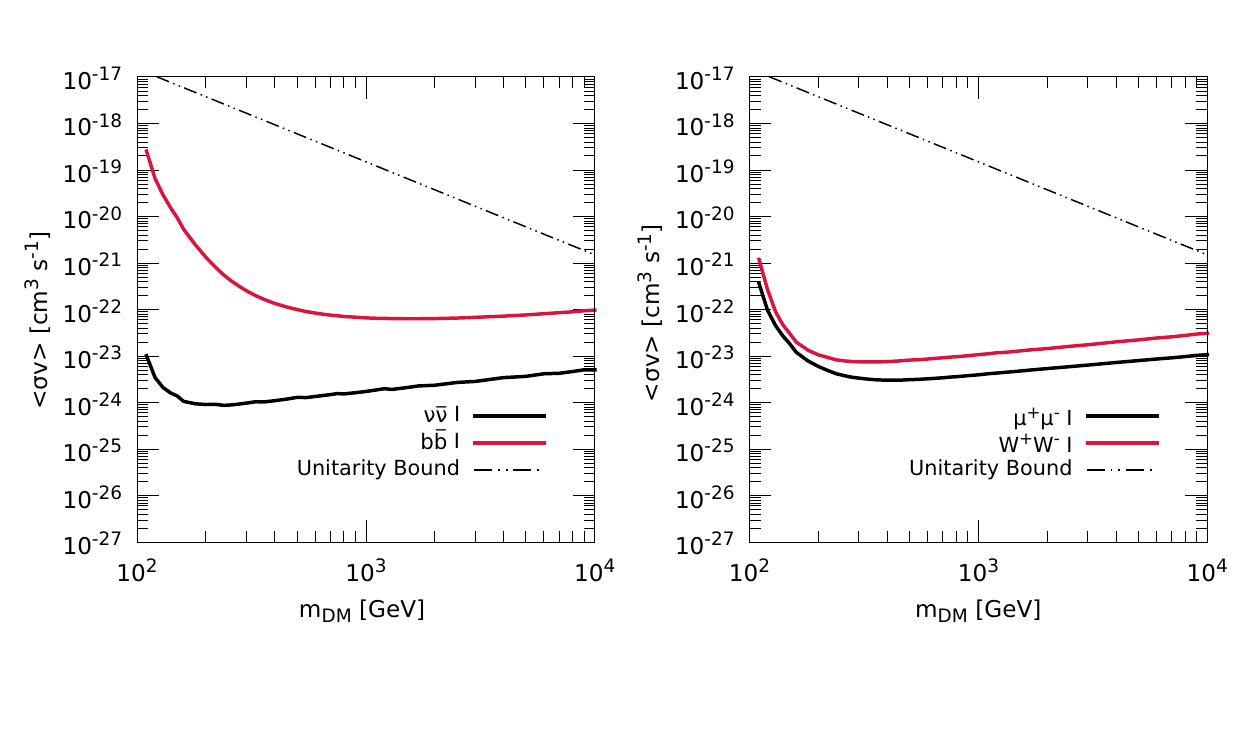} 
\vspace{-1.9cm} \\
\includegraphics[width=\textwidth]{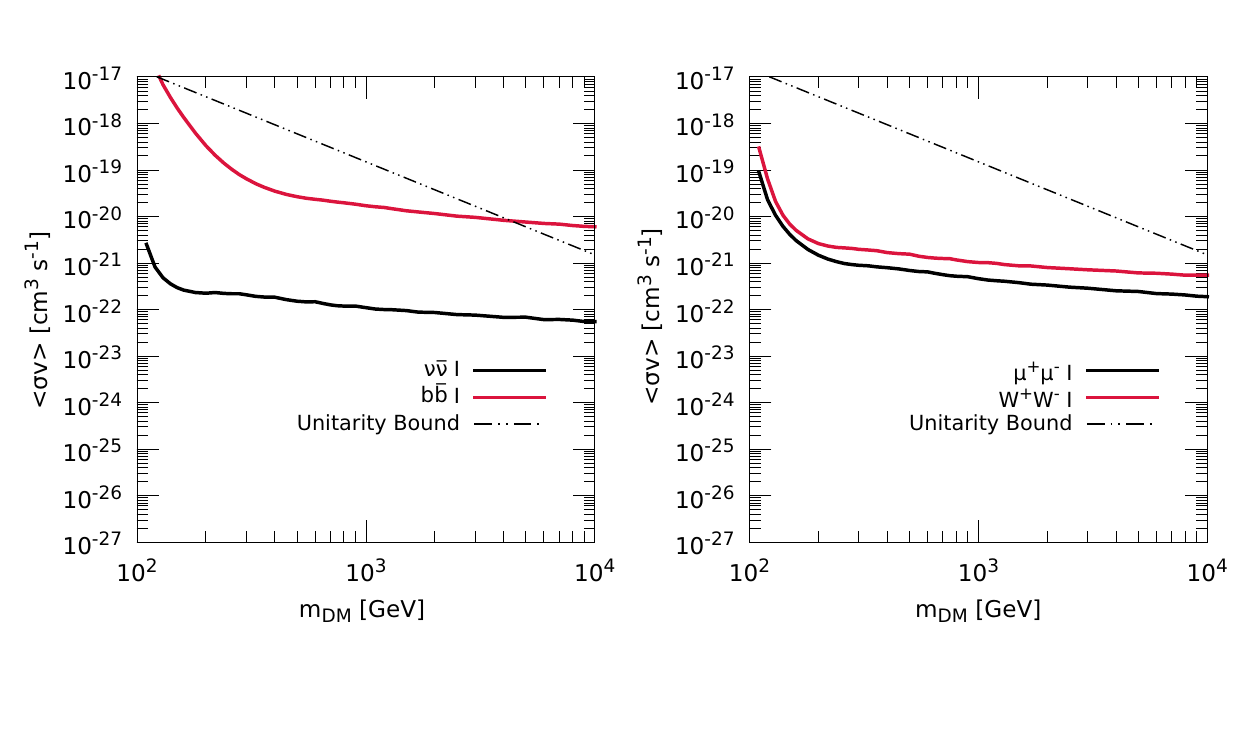}
\vspace{-1.5cm} 
\caption{Aggressive (upper panels) and conservative (lower panels) limits on the DM annihilation cross section at 90\%~C.L. with shower events after 10~years of data taking in a $1~{\rm km}^3$ neutrino telescope, for the combination I for the enhancement factor (solid lines). Left panels: Annihilations into $\nu\bar{\nu}$ (black lines) and into $b\bar{b}$ (red lines).  Right panels: Annihilations into $\mu^+\mu^-$ (black lines) and into $W^+W^-$ (red lines).  The unitarity limit~\cite{Griest:1989wd, Hui:2001wy} is shown as a double-dot-dashed line. 
}   
\label{fig:shlimits}
\end{figure}

In Fig.~\ref{fig:tglimits} we show the limits on the DM annihilation cross section in the mass range $m_{\rm DM} = [100~{\rm GeV}, 10~{\rm TeV}]$ for the case of through-going muon events, at 90\% C.L. The upper panels represent the results for the aggressive analysis and should be regarded as the most optimistic case for the setup we consider.  The conservative limits are depicted in the lower panels. For both analyses, the limits are shown for the enhancement factor I.  In the left panels, the limits for DM annihilations into $\nu\bar{\nu}$ (black lower line) and $b\bar{b}$ (red upper line) are shown, whereas in the right panels, we show the limits for DM annihilations into $\mu^+\mu^-$ (black lower line) and $W^+W^-$ (red upper line).  The limits for the $\mu^+\mu^-$ and $W^+W^-$ channels are similar to each other, for both produce relatively hard neutrino spectra, whereas the limits for the softest spectra of the four considered, DM annihilations into $b\bar{b}$, is the weakest.  Let us note that had we considered the most optimistic case, i.e., annihilations into $\nu\bar{\nu}$ (the hardest neutrino spectra) and enhancement factor IV, the aggressive limit would be about an order of magnitude above the thermal value for the annihilation cross section, for most of the mass range. As discussed above, this is not realistic, though. The uncertainties on the computation of the enhancement factor imply differences in the limits of about one order of magnitude, mainly coming from the differences in the way the results from N-body simulations are extrapolated to low masses and high redshifts.

In Fig.~\ref{fig:tglimits} we also depict the current IceCube and ANTARES 90\% C.L. limits of DM annihilations for track events with neutrinos in the direction of the galactic center.  Although these neutrinos are down-going neutrinos in IceCube, part of the detector can be used to veto the huge background of atmospheric muons and thus, events with the interaction vertex contained in the fiducial volume can be used for this type of searches.  On the other hand, ANTARES, being in the northern hemisphere, does not have this drawback and can actually use through-going muons, with a larger effective area. In Fig.~\ref{fig:tglimits} we show the limits with the 79-string  configuration in IceCube (contained vertex muons) after 320 days of data (black and red dotted lines)~\cite{Aartsen:2013mla} in both panels and, on the right panels, the ANTARES limits (through-going muons) for DM annihilations into $\tau^+\tau^-$ after about 1300 days of data (blue dotted lines)~\cite{Hernandez-Rey:2014ssa, Zornoza:2014cra}.  For the conservative analysis with the enhancement factor I, the limits are, in general, worse than those for neutrinos from the galactic center (at least for $m_{\rm DM}<10$~TeV).  However, with the aggressive analysis (or with the enhancement factor IV or with the combination of both), the limits on the DM annihilation cross section with cosmic neutrinos are expected to be better than those obtained with neutrinos in the detection of the galactic center, mainly for large DM masses and especially for IceCube.  In the case of ANTARES, only the most optimistic scenario with the most aggressive analysis would allow us to obtain more constraining limits with cosmic neutrinos than with neutrinos from the galactic center.  Finally, we also show (double-dot-dashed lines) the unitarity limit~\cite{Griest:1989wd, Hui:2001wy}.

In Fig.~\ref{fig:shlimits} we show the limits on the DM annihilation cross section in the mass range $m_{\rm DM} = [100~{\rm GeV}, 10~{\rm TeV}]$ for the case of shower events. We use the same notation of Fig.~\ref{fig:tglimits}.  Nevertheless, IceCube and ANTARES have not presented shower limits with neutrinos from the galactic center, so only our results and the unitarity limit are depicted in this figure. These limits are typically around an order of magnitude better than for the case of through-going muons.  As we already mentioned, although the effective size of the detector is larger for through-going muons produced in the surroundings of the detector, in the shower analyses, down-going (contained) events are also included and more importantly, the atmospheric neutrino background is about an order of magnitude lower for showers, as can be seen from Figs.~\ref{fig:tgspectra} and~\ref{fig:shspectra}.  This advantage of showers over tracks has already been noted in previous works~\cite{Mandal:2009yk, Lee:2012pz}, which we confirm.

\section{Conclusions}
\label{sec:conclusions}

Numerical CDM N-body simulations are a crucial tool for understanding structure formation in the Universe.  In particular, they allow the study the structural properties of individual halos and to predict the abundance of halos as a function of time and mass. Within these halos, DM annihilations take place. The neutrino luminosity from an individual halo depends on the total mass of the halo and on the distribution of DM particles within it. The combined neutrino emission from all halos at all redshifts produces an isotropic neutrino flux that could be searched for at neutrino telescopes~\cite{Beacom:2006tt, Murase:2012xs}.  However, the prediction of the neutrino flux suffers from uncertainties, mainly coming from the fact that these simulations only cover a limited range of halo masses and redshifts and extrapolations are necessary.

In this paper, we have estimated the reach of neutrino telescopes, such as IceCube or ANTARES, to detect cosmological DM annihilations after 10 years of data-taking for some selected final states.  We have discussed in detail the theoretical uncertainties in the limits associated to the modeling of the cosmological enhancement factor. Concretely, in Section~\ref{sec:nuflux}, we have considered three different parametrizations of the concentration parameter existing in the literature and have extrapolated the results of numerical N-body simulations to high redshifts and small halo masses (Tab.~\ref{tab:conc}) to compute the enhancement factor for the DM annihilation signal due to individual halos (Fig.~\ref{fig:xiM2}). In addition, we have considered three parametrizations of the  halo mass function (the function which describes the abundance of halos as a function of mass and redshift) also derived from N-body simulations (Tab.~\ref{tab:f} and Fig.~\ref{fig:dndM}). We have then calculated the cosmological enhancement factor for DM annihilations for various combinations of the concentration parameter and the halo mass function, in order to assess the theoretical uncertainty on this quantity (Tab.~\ref{tab:combinations} and Fig.~\ref{fig:xi2}), which translates into an uncertainty of about an order of magnitude in the cosmic neutrino flux. However, as have been recently found~\cite{Ishiyama:2010es, Anderhalden:2013wd, Ishiyama:2014uoa}, extrapolations of simple power laws for the concentration--mass relation to very low halo masses are highly disfavored~\cite{Sanchez-Conde:2013yxa, Ludlow:2013vxa}, so the combination IV overpredicts the expected rates. Let us also note that we have not included the contribution from substructure, which could enhance the flux by a factor of a few~\cite{Kamionkowski:2010mi, SanchezConde:2011ap, Fornasa:2012gu, Sanchez-Conde:2013yxa}, or even by an order of magnitude more~\cite{Springel:2008cc, Pinzke:2011ek, Gao:2011rf}. In this regard, our results are conservative.

In Section~\ref{sec:event}, we have described the calculation of the event spectra for the two type of events in neutrino telescopes we have considered in this work: through-going muon events and showers. Examples of these spectra, compared to that of the atmospheric neutrino background, are depicted in Figs.~\ref{fig:tgspectra} and~\ref{fig:shspectra}.  

Finally, using a simplified modeling of the detector, in Section~\ref{sec:bounds} we have estimated the limits on the annihilation cross section after 10 years of data-taking in a Gton-scale neutrino telescope for several exemplary final states, assuming that no signal is observed.  We have considered two different analyses: an aggressive analysis, for which we have only assumed statistical errors; and a conservative analysis, for which we have included systematic errors and restricted the DM signal to be smaller than the background in all energy bins.  In both cases, our ignorance of the structural properties of the individual halos and on the abundance of low-mass halos translates into an uncertainty band in the limits of approximately an order of magnitude, when considering the enhancement factors I, II and III (see Fig.~\ref{fig:xi2}). Note that we do not include the combination IV in this estimate. By doing so, one would overestimate the theoretical uncertainty on the signal from the smooth distribution to be about two orders of magnitude.  

Our results are shown in Figs.~\ref{fig:tglimits} and~\ref{fig:shlimits}.  As has already been pointed out in the context of DM annihilations in the galactic halo~\cite{Mandal:2009yk, Lee:2012pz}, we stress that the sensitivity of neutrino telescopes to the DM annihilation cross section with showers is potentially better than with through-going muon events, because the atmospheric electron and tau neutrino flux is much smaller than the muon neutrino flux at energies above 100~GeV and hence, the signal-to-noise ratio is larger for the case of showers.     

Moreover, we have noticed that, despite the uncertainties, for neutrino telescopes not directly observing the galactic center, as is the case of IceCube, the sensitivity to DM annihilations with observations of the cosmic diffuse neutrino signal could be competitive with observations of neutrinos in the direction of the galactic center (Fig.~\ref{fig:cosmo_vs_GC}).

\acknowledgments
We would like to thank J.~Hern\'andez-Rey and J.~D.~Zornoza for discussions about ANTARES and for providing us with the current limits with neutrinos in the direction of the galactic center.  We also thank M.~A.~S\'anchez-Conde for a careful reading of the manuscript.  Finally, we would like to thank the anonymous referee for useful comments and suggestions. AM is supported by the Funda\c{c}\~ao para a Ci\^encia e a Tecnologia (FCT) of Portugal.  AI is partially supported by the DFG cluster of excellence ``Origin and Structure of the Universe''.  SPR is supported by a Ram\'on y Cajal contract, by the Spanish MINECO under grant FPA2011-23596 and by GVPROMETEOII/2014/049.  AM and SPR are also partially supported by the Portuguese FCT through the projects PTDC/FIS-NUC/0548/2012 and CFTP-FCT Unit 777 (PEst-OE/FIS/UI0777/2013), which are partially funded through POCTI (FEDER), and SPR is also supported by the European Union FP7 ITN INVISIBLES (Marie Curie Actions, PITN-GA-2011-289442).

\pagebreak

\small
\bibliographystyle{JHEP}
\bibliography{cosmicnusbiblio}

\end{document}